\documentclass[modern]{aastex631}

\usepackage{amsmath}
\usepackage{soul}

\newcommand{\mgfe}[0]{[{\rm Mg/Fe}]}

\newcommand{\mgh}{[{\rm Mg}/{\rm H}]}
\newcommand{\feh}[0]{[{\rm Fe/H}]} 
 
\newcommand{\xh}{[{\rm X}/{\rm H}]}

\newcommand{\logg}{\log(g)}
\newcommand{\teff}{T_{\rm eff}}
\newcommand{\kpc}{\rm \; kpc}

\newcommand{\Rg}{R_{\rm guide}}
\newcommand{\Zm}{Z_{\rm max}}

\newcommand{\N}{67315}

\shorttitle{Many elements matter}
\shortauthors{griffith and hogg}
\renewcommand{\paragraph}[1]{\bigskip\par\noindent{\textbf{#1}}~---}
\sloppypar

\graphicspath{{./}{}}

\begin{document}

\title{Many elements matter: Detailed abundance patterns reveal star-formation and enrichment differences among Milky Way structural components}

\correspondingauthor{Emily J. Griffith}
\email{Emily.Griffith-1@colorado.edu}

\author[0000-0001-9345-9977]{Emily J. Griffith}
\altaffiliation{NSF Astronomy and Astrophysics Postdoctoral Fellow}
\affiliation{Center for Astrophysics and Space Astronomy, Department of Astrophysical and Planetary Sciences, University  of Colorado, 389~UCB, Boulder,~CO 80309-0389, USA}

\author[0000-0003-2866-9403]{David W. Hogg}
\affiliation{Center for Cosmology and Particle Physics, Department of Physics, New York University, 726~Broadway, New~York,~NY 10003, USA}
\affiliation{Max-Planck-Institut f{\"u}r Astronomie, K{\"o}nigstuhl 17, D-69117 Heidelberg, Germany}
\affiliation{Center for Computational Astrophysics, Flatiron Institute, 162 Fifth Avenue, New York, NY 10010, USA}

\author[0000-0001-5388-0994]{Sten Hasselquist}
\affiliation{Space Telescope Science Institute, 3700 San Martin Drive, Baltimore, MD 21218, USA}

\author[0000-0002-6534-8783]{James W. Johnson}
\affiliation{Carnegie Science Observatories, 813 Santa Barbara St., Pasadena, CA 91101, USA}

\author[0000-0003-0872-7098]{Adrian Price-Whelan}
\affiliation{Center for Computational Astrophysics, Flatiron Institute, 162 Fifth Avenue, New York, NY 10010, USA}

\author[0000-0001-8208-9755]{Tawny Sit}
\affiliation{The Department of Astronomy and Center of Cosmology and AstroParticle Physics, The Ohio State University, Columbus, OH 43210, USA}

\author[0000-0003-4761-9305]{Alexander Stone-Martinez }
\affiliation{Department of Astronomy, New Mexico State University, P.O.Box 30001, MSC 4500, Las Cruces, NM, 88033, USA}

\author[0000-0001-7775-7261]{David H. Weinberg}
\affiliation{The Department of Astronomy and Center of Cosmology and AstroParticle Physics, The Ohio State University, Columbus, OH 43210, USA}

\begin{abstract}\noindent 
Many nucleosynthetic channels create the elements, but two-parameter models characterized by $\alpha$ and Fe nonetheless predict stellar abundances in the Galactic disk to accuracies of 0.02 to 0.05~dex for most measured elements, near the level of current abundance uncertainties.
It is difficult to make individual measurements more precise than this to investigate lower-amplitude nucleosynthetic effects, but population studies of mean abundance patterns can reveal more subtle abundance differences.
Here we look at the detailed abundances for $\N$ stars from APOGEE DR17, but in the form of abundance residuals away from a best-fit two-parameter, data-driven nucleosynthetic model.
We find that these residuals show complex structures with respect to age, guiding radius, and vertical action that are not random and are also not strongly correlated with sources of systematic error such as $\logg$, $\teff$, and radial velocity.
The residual patterns, especially in Na, C+N, Ni, Mn, and Ce, trace kinematic structures in the Milky Way, such as the inner disk, thick disk, and flared outer disk.
A principal component analysis suggests that most of the observed structure is low-dimensional and can be explained by a few eigenvectors.
We find that some, but not all, of the effects in the low-$\alpha$ disk can be explained by dilution with fresh gas, so that abundance ratios resemble those of stars with higher metallicity.
The patterns and maps we provide could be combined with accurate forward models of nucleosynthesis, star formation, and gas infall to provide a more detailed picture of star and element formation in different Milky Way components.

\end{abstract}


\section{Intro}\label{sec:intro}

Stellar abundances as observed by large, medium resolution surveys like GALAH \citep[GALactic Archaeology with HERMES;][]{desilva2015} and APOGEE/MWM \citep[Apache Point Observatory Galactic Evolution Experiment, Milky Way Mapper;][]{majewski2017, kollmeier2019} can successfully be predicted by two-parameter models of $\alpha$ and Fe or age and radius \citep[e.g.,][hereafter G24]{frankel2018, weinberg2022, ness2022, ratcliffe2023, griffith2024}. Two-parameter models conditioned on $\alpha$ and Fe aim to capture the typical enrichment from core-collapse supernovae (CCSN) and Type-Ia supernovae (SNIa). These simple models can predict $10$ to $15$ [X/H] abundances to a precision of 0.02 to 0.05 dex, near the reported abundance errors within these spectroscopic surveys.

To this level of precision, a single two-parameter model can successfully predict stellar abundances of disk stars at all Galactic radii \citep{weinberg2019, griffith2021a}. 
This universality is surprising, given the changing metallicity distribution function (MDF) of stars across the Galactic disk \citep[e.g.,][]{bensby2013, hayden2015}. 
The MDF is sensitive to aspects of chemical evolution such as star formation efficiency (SFE), inflows, outflows, and initial mass function (IMF) \citep[e.g.,][]{andrews2017}. 
Its spatial dependence suggests that some aspects of chemical evolution change with radius and height and that these histories may be encoded in the observed abundance distributions. 

Two-parameter models, however, cannot perfectly predict all stellar abundances. 
Residuals away from the model predictions are powerful tools to search for enrichment that does not depend on integrated CCSN and SNIa yields alone. 
At large amplitudes (tenths of dex), two-parameter model residuals can identify stars or stellar populations with atypical enrichment, including dwarf galaxies and globular clusters \citep{weinberg2022, sit2024}. 
At low amplitudes (hundredths of dex), cross-element residual correlations robustly reveal the signature of a third asymptotic giant branch star (AGB)-like nucleosynthetic process, when many neutron-capture elements are available \citep{griffith2022}. Certain low-amplitude abundance residuals also correlate with parameters such as radius and age \citep{ness2022, ratcliffe2023}. 
These residual abundance correlations show that, at small scales, a two-process model cannot fully describe the abundance distributions of the MW. Our understanding of the cause of these residuals is not yet complete. In the context of multi-parameter nucleosynthetic models, we have yet to explore what elements beyond Mg and Fe (and occasionally neutron capture elements) can teach us about the MW disk.

Simultaneously, principal component analysis (PCA) and latent variable space analysis of APOGEE and GALAH abundances find $5-10$ components are needed to explain the abundance distributions observed in the MW \citep[e.g.,][]{price2018, casey2019, ratcliffe2020, ratcliffe2022}, similar to the earlier PCA-based conclusions of \citet{ting2012}. In an analysis of APOGEE data, \citet{ting2022} find that it is necessary to condition on seven elements, not just two, to reduce residual cross-element correlations to a level expected from observational uncertainty alone. While cross-element residual correlations are sensitive to systematic abundance trends with stellar parameters, searching for population-level residual abundance correlations with less-biased parameters, such as orbits or actions, may enable the detection of meaningful residual trends within large stellar populations below the reported level of abundance uncertainty. 

In this article, we explore trends between residuals away from a two-parameter model, (KPM; G24) with age, guiding radius, and maximum height above the midplane. 
If stellar abundances depend not only on Mg and Fe, but also on the combination of processes that enriches the gas that ultimately forms the stars, then we expect to see residuals away from the two-parameter model that correlate with the kinematic structures of the disk. Using a sample of $\logg$ calibrated stellar abundances for 15 elements from APOGEE DR17 (Section~\ref{sec:data}), we fit a $K=2$ process model with KPM (Section~\ref{sec:fitting_kpm}) and calculate abundance residuals away from the model predictions. We map these abundance residuals in orbital and age spaces (Section~\ref{sec:resids}) and quantify the dimensionality of the observed structure (Section~\ref{subsec:PCA}). Finally, we explore potential causes of the observed abundance structure (Section~\ref{sec:causes_resids}) and discuss the implications for decoding the Galaxy's chemical evolution history (Section~\ref{sec:discussion}).

\section{Data Sample}\label{sec:data}

We employ stellar abundances from APOGEE DR17 \citep{abdurrouf2022}, part of the SDSS-IV survey \citep{blanton2017, majewski2017} in our analysis. APOGEE DR17 reports data for 657,135 unique stars observed with twin spectrographs on the 2.5m Sloan Foundation telescope \citep{gunn2006, wilson2019} at Apache Point Observatory in New Mexico and the 2.5m du Pont Telescope \citep{bowen1973} at the Las Campanas Observatory in Chile. These high-resolution (R $\sim 22,500$) infrared spectrographs observe targets that span the Galactic disk, inner Galaxy, halo, and nearby satellites. \citet{zasowski2013, zasowski2017}, \citet{beaton2021} and \citet{santana2021} describe the survey targeting in detail. Stellar spectra are reduced and calibrated with the APOGEE data processing pipeline \citep{nidever2015} after which stellar parameters and abundances are derived with the APOGEE Stellar Parameter and Chemical Abundance Pipeline \citep[ASPCAP;][]{holtzman2015, garcia2016}. \citet[][DR16]{jonsson2020} and Holtzman et al. (in prep, DR17) provide additional details on the spectral analysis process as well as the precision of the abundance measurements. 

APOGEE DR17 reports stellar abundances for 20 species: C, C I, N, O, Na, Mg, Al, Si, S, K, Ca, Ti I, Ti II, V, Cr, Mn, Fe, Co, Ni, and Ce. NLTE corrections are implemented for Na, Mg, K, and Ca \citep{osorio2020, hubeny2021}. We exclude Ti I, Ti II,  P, and V from our analysis due to their lack of precision, scatter, and abundance artifacts, as described in \citet{jonsson2020}. As in G24, we are interested in fitting and predicting the birth abundances of stars. While the surface abundance of most elements remains unchanged over a star's lifetime, the CNO process and dredge-up events alter the C and N surface abundance while the sum of C$+$N remains unchanged \citep{salaris2015, vincenzo2021b}. We consider C$+$N as an element, where [C$+$N/H] is
\begin{equation}
    [\text{C+N}/\text{H}] = \log_{10}(10^{\text{[C/H]}+8.39} + 10^{\text{[N/H]}+7.78}) - \log_{10}(10^{8.39} + 10^{7.78}),
\end{equation}
using solar abundances of 8.39 and 7.78 for C and N, respectively \citep{grevesse2007}. We adopt the [C/Fe] error as the error on  [C$+$N/Fe] as C dominates the abundance ratio.

In addition to the main survey products, we are interested in analyzing the residual abundance correlation with orbital parameters and ages. We compute Galactic orbital parameters and actions using \texttt{gala} \citep{Price-Whelan:2017}. We use distance estimates from \texttt{STARHORSE} \citep{starhorse}, astrometry from Gaia DR3 \citep{gaia2023}, and radial velocities from APOGEE DR17 and transform to Galactocentric positions and velocities using Astropy \citep{astropy2013, astropy2018, astropy2022}. We adopt a sun–Galactic center distance of $R_0 = 8.275~\textrm{kpc}$ \citep{Gravity:2021}, a Solar height above the midplane of 20.8 pc \citep{Bennett:2019}, and a total Solar velocity with respect to the Galactic center of $\boldsymbol{v}_\odot = (8.4, 251.8, 8.4)~\textrm{km}~\textrm{s}^{-1}$ calculated from the radial velocity and proper motion of Sgr A* \citep{Reid:2020, Gravity:2021, Hunt:2022}. We use the ``O2GF'' method for computing estimates of the orbital actions \citep{Sanders:2014, Sanders:2016} using the \texttt{MilkyWayPotential2022} mass model implemented in \texttt{gala}.

For stellar ages, we adopt values from the \texttt{distmass} catalog described in \citet{stone2024}. Specifically, we adopt ages from the \texttt{distmass} uncorrected Sharma model, where stellar masses are estimated through a neural network trained on the ASPCAP parameters, including [C/N], of stars with asteroseismic masses from APOKASC 3 \citep{sharma2016, pinsonnealt2024}. Ages are then inferred from MIST isochrones \citep{choi2016}. Due to the parameter space of the APOKASC training data, ages are only available for stars with $\logg > 1$ and [Fe/H] $> -0.5$.

\subsection{Stellar Abundance Calibrations}\label{subsec:calibration}

For the last decade, we have studied the MW's stellar abundances as a function of position across much of the Galactic disk with APOGEE \citep[e.g.,][]{hayden2014, hayden2015, weinberg2019, griffith2021a, weinberg2022}. 
Unfortunately, stellar abundance determination is imperfect, and uncertainties in model atmospheres and analysis pipelines can cause abundance systematics on the scale of $\sim 0.05 dex$ that correlate with stellar parameters, such as $\logg$ and $\teff$. These systematics must be accounted for in order to robustly compare stellar abundances between two populations with different $\logg$ distributions, such as the MW bulge and the disk \citep{griffith2021a}, since more luminous stars are found at larger distances from the Sun. If not corrected, these systematics will cause $\logg$, and thus radially dependent abundance trends, as well as correlated residuals from a multi-parameter nucleosynthetic model.

To address this issue, \citet[][hereafter S24]{sit2024} derive an empirical calibration to correct for $\logg$ dependent abundance trends in evolved stars within the APOGEE DR17 catalog. For this correction, they assume that the median [X/Mg] vs. [Mg/H] abundance trends should be independent of $\logg$. From a smaller, high signal-to-noise training set, they derive metallicity and $\logg$ dependent offsets for each element to bring median abundance trends into agreement across $\logg$ bins from 0 to 3.5. They also calculate zero-point offsets to ensure stars of solar [Mg/H] have solar [X/Mg]. S24 then apply this calibration, derived from a training set of 151,564 stars, to 310,427 evolved stars in APOGEE with ASPCAP abundances. This calibration shifts all stellar abundances to what they would be observed as at $\logg=1.75$. 

In this paper, we adopt the calibrated ASPCAP stellar abundance catalog from S24 as our parent sample. The S24 catalog is restricted to stars with: 
\begin{enumerate}
    \item $-0.75 \leq \mgh \leq 0.45$
    \item $0 \leq \logg \leq 3.5$
    \item $3000 \leq \teff \leq 5500$
    \item S/N $> 80$
\end{enumerate}
We further restrict our sample to stars with:
\begin{enumerate}
    \item \texttt{EXTRATARG}=0 (main survey sample with no duplicates)
    \item $Z_{\rm max} < 2 \kpc$
    \item $R_{\rm guide} < 16 \kpc$
    \item $\logg < 2.3$ (to exclude the red clump)
\end{enumerate}

These cuts result in a stellar sample of $67,315$ stars that span the Galactic disk, 44,655 of which have ages from \citet{stone2024}. Of these totals, 46,463 (33,285 with ages) are part of the chemical high-Ia (low-$\alpha$) population and 20,852 (11,370 with ages) are part of the chemical low-Ia (high-$\alpha$) population\footnote{Traditionally referred to as the low-$\alpha$ and high-$\alpha$ populations, we adopt the language coined in \citet{griffith2019} of high-Ia and low-Ia, respectively, alluding to the enrichment origins of the abundance differences.}, where the low-Ia population is defined as
\begin{equation}\label{eq:lowIa}
\begin{cases}
\mgfe > 0.12 - 0.13 \, \feh,    & \feh<0 \cr
\mgfe > 0.12,               & \feh>0, \cr
\end{cases}
\end{equation}
using the raw APOGEE abundances, as in previous works \citep[e.g.,][S24]{weinberg2022}.

\begin{figure*}[htb!]
    \centering
    \includegraphics[width=1\linewidth]{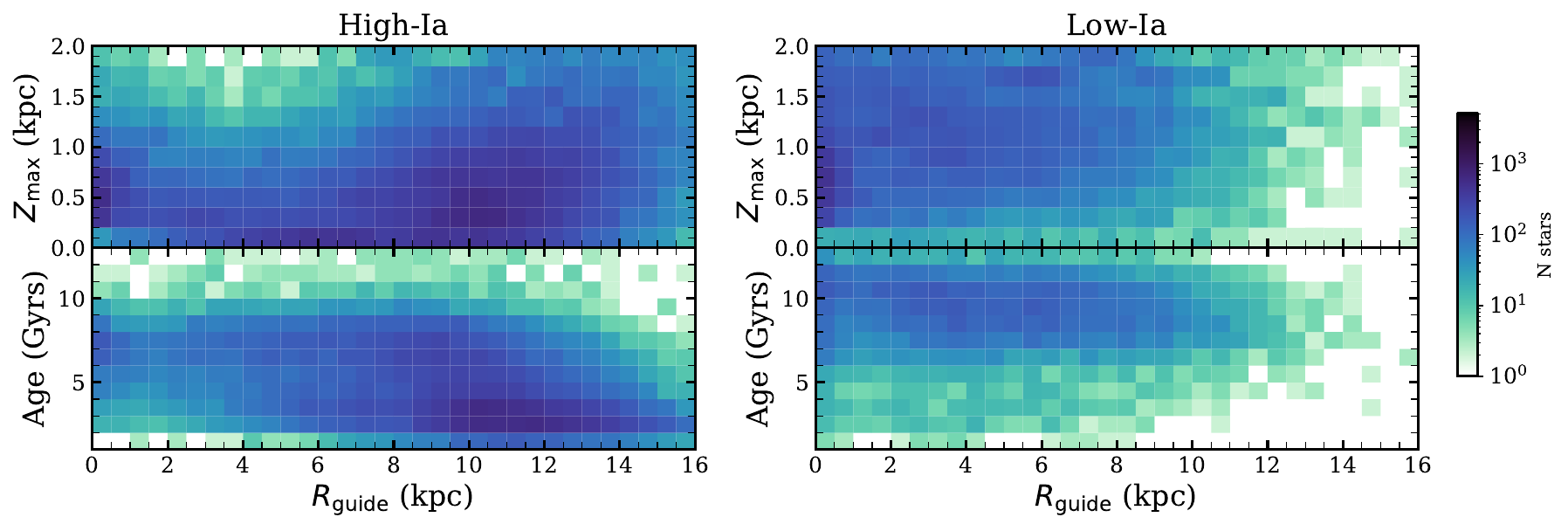}
    \caption{Distribution of our high-Ia (left) and low-Ia (right) samples in $\Zm$ vs. $\Rg$ (top) and age vs. $\Rg$ (bottom). }
    \label{fig:density_maps.pdf}
\end{figure*}

In Figure~\ref{fig:density_maps.pdf} we plot the distributions of our low-Ia and high-Ia samples in $\Zm$ vs. $\Rg$  and age vs. $\Rg$---the dimensions we will focus on in this paper.
We choose to work in orbital invariant quantities $\Zm$ and $\Rg$ rather than cylindrical $R$ and $Z$ positions to minimize potential blurring of important trends from the variations of stellar positions as stars travel on their orbits;
for each star, $\Zm$ is the maximum vertical position $Z$ to which that star on its current orbit will rise; and $\Rg$ is the radius corresponding to a circular, in-plane orbit with the same angular velocity as the star.
These invariants are like invariant actions, but they are closer to the observables; they have intuitive explanations and can be plotted like maps of the Galaxy.
As expected in these plots, we can see the ``thin'' and ``thick'' disks. The high-Ia stars are young (age $\lesssim 8$ Gyrs) and lie on orbits that span $\Rg$, with low $\Zm$ in the inner Galaxy and larger $\Zm$ in the outer Galaxy. Conversely, the low-Ia stars are predominantly old (age $\gtrsim 7$ Gyrs) and lie on orbits with higher $\Zm$ and $\Rg$ less than 10. The high-Ia disk is well sampled within the parameter limits that we have chosen, and the low-Ia disk is well sampled at $\Rg < 10$.

\section{The Fiducial K-Process Model} \label{sec:fitting_kpm}

In this paper, we employ KPM\footnote{\url{https://github.com/13emilygriffith/KProcessModel}}, developed by G24 and based on the two-parameter model of \citet{weinberg2019}. This multi-parameter nucleosynthetic model is data-driven and can be flexibly implemented by the user. It assumes that all stellar abundances can be described by the sum of $K$ nucleosynthetic processes, where each element, $j$, has $K$ metallicity-dependent process vectors, $q_{k,j}^Z$, and each star, $i$ has $K$ process amplitudes, $A_{i}^k$. Under KPM, the abundance of element $j$ relative to H in star $i$ ($m_{ij}$) is expressed as
\begin{equation}\label{eq:kpm1}
        m_{ij} = \log_{10} \sum_{n=1}^K A_{i}^n \, q_{n,j}^Z.
\end{equation}
The observed stellar abundance is thus
\begin{equation}
    [\text{X}_j/\text{H}]_i = m_{ij} + \text{noise},
\end{equation}
where ``noise'' represents sources of intrinsic scatter and observational errors that are not accounted for in our model.

As in G24, KPM adopts a set of assumptions to make the model interpretable, break rotational degeneracies, and find the best model parameters. We summarize those assumptions below, noting changes and improvements from the G24 version.
\begin{enumerate}
    \item The abundances of all elements on the periodic table can be explained by a combination of $K$ nucleosynthetic processes. We assume $K=2$ in the fiducial model.
    \item The (X/H) abundances of a star can be described by a linear combination of these $K$ processes, as in Equation~\ref{eq:kpm1}.
    \item All model process amplitudes and process vector components are positive. 
    \item Each process is fixed to some element or set of elements with specified process vectors. In the fiducial model, we fix the first process to Mg and the second process to Fe, chosen to make the processes interpretable as CCSN and SNIa-like sources such that
    \begin{equation}\label{eq:qcc_mg_solar}
    q_{{\rm 1, Mg}}^{\,Z} = 1, \quad q_{2,{\rm Mg}}^{\,Z} = 0
    \end{equation}
    \begin{equation}\label{eq:qcc_fe_solar}
    q_{{\rm 1, Fe}}^{\,Z} = 0.4, \quad q_{2,{\rm Fe}}^{\,Z} = 0.6
    \end{equation}
    \item The process vector components for all elements except those fixed to each process above are allowed to be metallicity dependent, where [Mg/H] is the default metallicity parameter. This is implemented by fitting the coefficients of $J-1$ sins and cosines away from some constant value and allows for complex yet smooth metallicity dependence in the process vector components, an update from G24. In the fiducial model, we set $J=11$. Increasing $J$ further does not significantly improve the quality of the fits.
    \item The APOGEE abundances and uncertainties can be interpreted in the context of this model, though the derived process parameters may be impacted by abundance systematics such as those described in Section~\ref{sec:data}. 
    \item The log likelihood function underlying our model is given by the objective 
    \begin{equation}\label{eq:chi2}
    \chi^2 = \sum_{ij}\frac{1}{\sigma_{ij}^2} \, (\xh_{ij} - m_{ij})^2 ~.
    \end{equation}
    where $1/\sigma_{ij}^2$ is the robust inverse variance on value $ij$:
    \begin{equation}\label{eq:inflate_ivar}
    \frac{1}{\sigma_{ij}^2} = \frac{Q^2/\sigma_{{\rm obs},ij}^2}{Q^2 + (\xh_{ij} - m_{ij})^2 /\sigma_{{\rm obs},ij}^2 }.    
    \end{equation}
    As discussed in G24, $Q$ is a softening parameter that inflates the observed abundance errors. The model is largely insensitive to the choice of $Q$.
    \item The likelihood function is optimized on a set of stellar abundances using a Gauss-Newton nonlinear least-squares minimization algorithm in \texttt{jaxopt}. KPM iteratively optimizes the process vector components and the process amplitudes over 48 steps. Taking fewer steps does not significantly change the model fits, as the $\chi^2$ is optimized quickly.
\end{enumerate}
See G24 for a more in-depth discussion of the model's assumptions, performance, and limitations.

We fit KPM to our sample of $\N$ stars with a $K=2$ model. In this case, Equation~\ref{eq:kpm1} becomes
\begin{equation}\label{eq:kpm_fid}
        m_{ij} = \log_{10} ( A_{i}^1 \, q_{1,j}^Z +  A_{i}^2 \, q_{2,j}^Z)
\end{equation}
The process vectors ($q_{1,j}^Z$ and $q_{2,j}^Z$) are fit to the abundance distributions of each element and the process amplitudes ($A_{i}^1$ and $A_{i}^2$) are fit to the Mg and Fe abundances of each star. Fitting the amplitudes to Mg and Fe minimizes the anti-correlations in the residuals of like elements (e.g., O and Mg), that appear when more elements are included in the fits, and it simplifies the interpretation of our results because only two elements enter the predicted abundances. This choice also results in the perfect prediction of the Mg and Fe abundances. The optimized process vector components and process amplitudes are very similar to those from G24 and S24, so we do not show or discuss them here.

\section{Abundance Residuals}\label{sec:resids}

From the KPM best fit parameters, we calculate the $K=2$ predicted abundances according to Equation~\ref{eq:kpm1}. These values represent the abundances each star would have if the model assumptions are correct--that the elements are created by two nucleosynthetic processes, and there are no observational systematics. In practice, the observed abundances will deviate from the predicted abundances due to statistical and systematic abundance errors including imperfect spectral modeling \citep[e.g.,][S24]{griffith2021a}, as well as unaccounted for physics, such as additional nucleosynthetic sources \citep[e.g.,][]{griffith2022}, IMF variations, metallicity-dependent yields, and/or other impacts of chemical evolution. If detected, these latter deviations from the KPM predictions could be a powerful tool to study nucleosynthesis and chemical enrichment at the 0.02 to 0.1 dex scale. 

In the following subsections, we will show maps of the average abundance residuals in $\Zm$ vs. $\Rg$  and age vs. $\Rg$ and describe the observed residual structures. In Section~\ref{subsec:PCA}, we will quantify the dimensionality of the residuals. As shown in Figure~\ref{fig:density_maps.pdf}, the high-Ia and low-Ia division from Equation~\ref{eq:lowIa} roughly separates stars into the young thin disk and the old thick disk, though this chemical separation does not perfectly separate the kinematic components \citep{hayden2017}. These two populations have distinct evolutionary histories \citep[e.g.,][]{chiappini1997, chandra2024, spitoni2024}, so we will study them separately. 

\subsection{Residual Structure in Maximum Height vs. Guiding Radius}\label{subsec:RvsZ}

\begin{figure}
    \centering
    \includegraphics[width=1\linewidth]{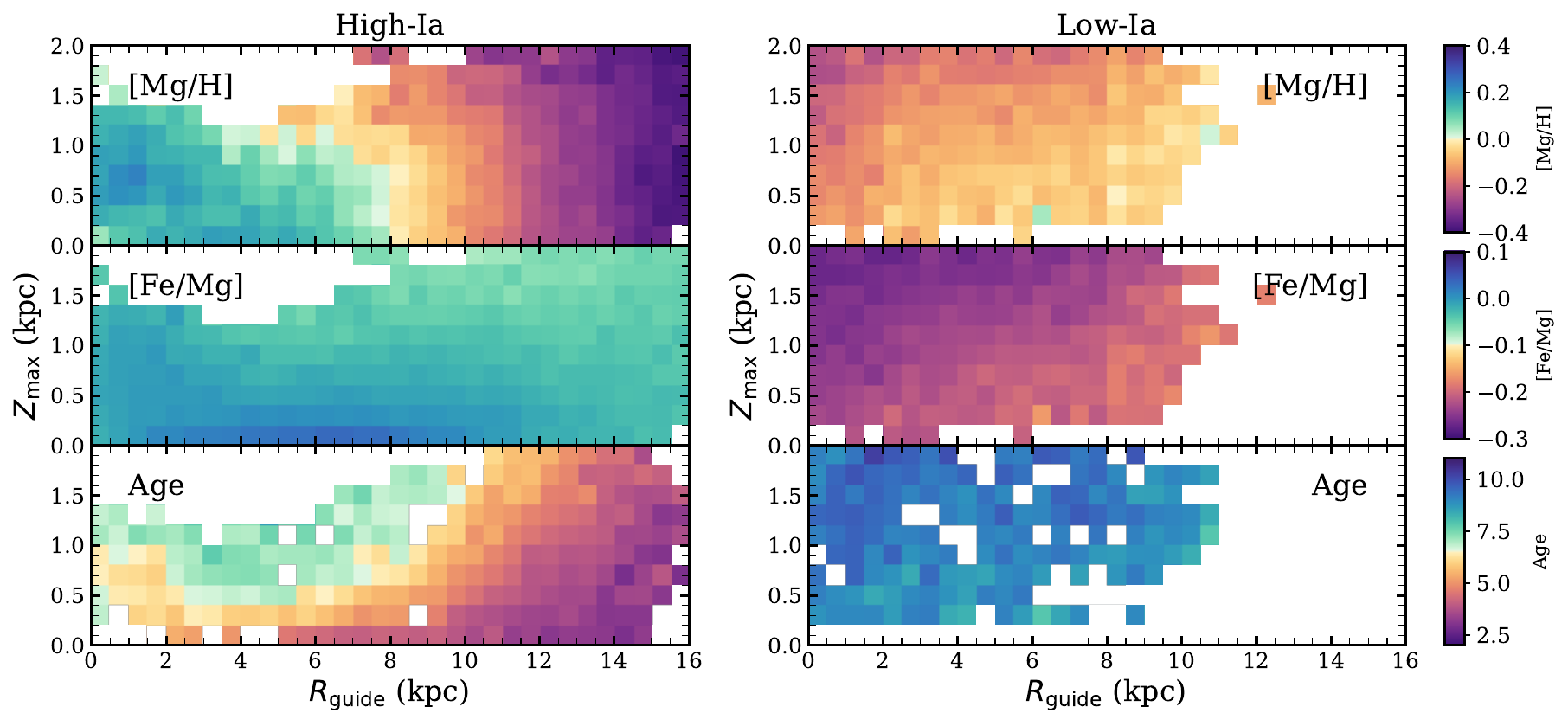}
    \caption{Maps of the average [Mg/H] (top), [Fe/Mg] (middle), and age (bottom) for the high-Ia (left) and low-Ia (right) populations in $\Zm$ vs. $\Rg$. Bins with less than 20 stars have been excluded.}
    \label{fig:MgFe_map_Z_R}
\end{figure}

In this Section, we investigate the residual structure in maps of $\Zm$ vs. $\Rg$. We plot the average [Mg/H], [Fe/Mg], and age in this space in Figure~\ref{fig:MgFe_map_Z_R}. Here, we see the metallicity gradient across the high-Ia (left) and low-Ia (right) populations in [Mg/H]: the high-Ia stars are metal-rich at small $\Rg$ to metal-poor at large $\Rg$ and the low-Ia stars are metal-rich at low $\Zm$ and metal-poor at high $\Zm$ and low $\Rg$. In [Fe/Mg], we see the expected differences in the high-Ia and low-Ia populations, which are high-[Fe/Mg] and low-[Fe/Mg], respectively, by construction. Some disk structure is visible in the [Fe/Mg] gradient for the high-Ia population, where the inner disk at low $\Zm$ stands out with higher [Fe/Mg]. As we are fitting the KPM amplitudes to Mg and Fe, the observed [Mg/H] and [Fe/Mg] trends will be perfectly reproduced by the model. In the average age map, we also see a clear trend, with young stars flaring to larger heights at large ($\Rg\gtrsim8$) kpc. These abundance and age trends have been well characterized in the literature \citep[e.g.,][]{hayden2015, bovy2016, cunha2016, martig2016}, but we show them here for reference.

\begin{figure*}
    \centering
    \includegraphics[width=1\linewidth]{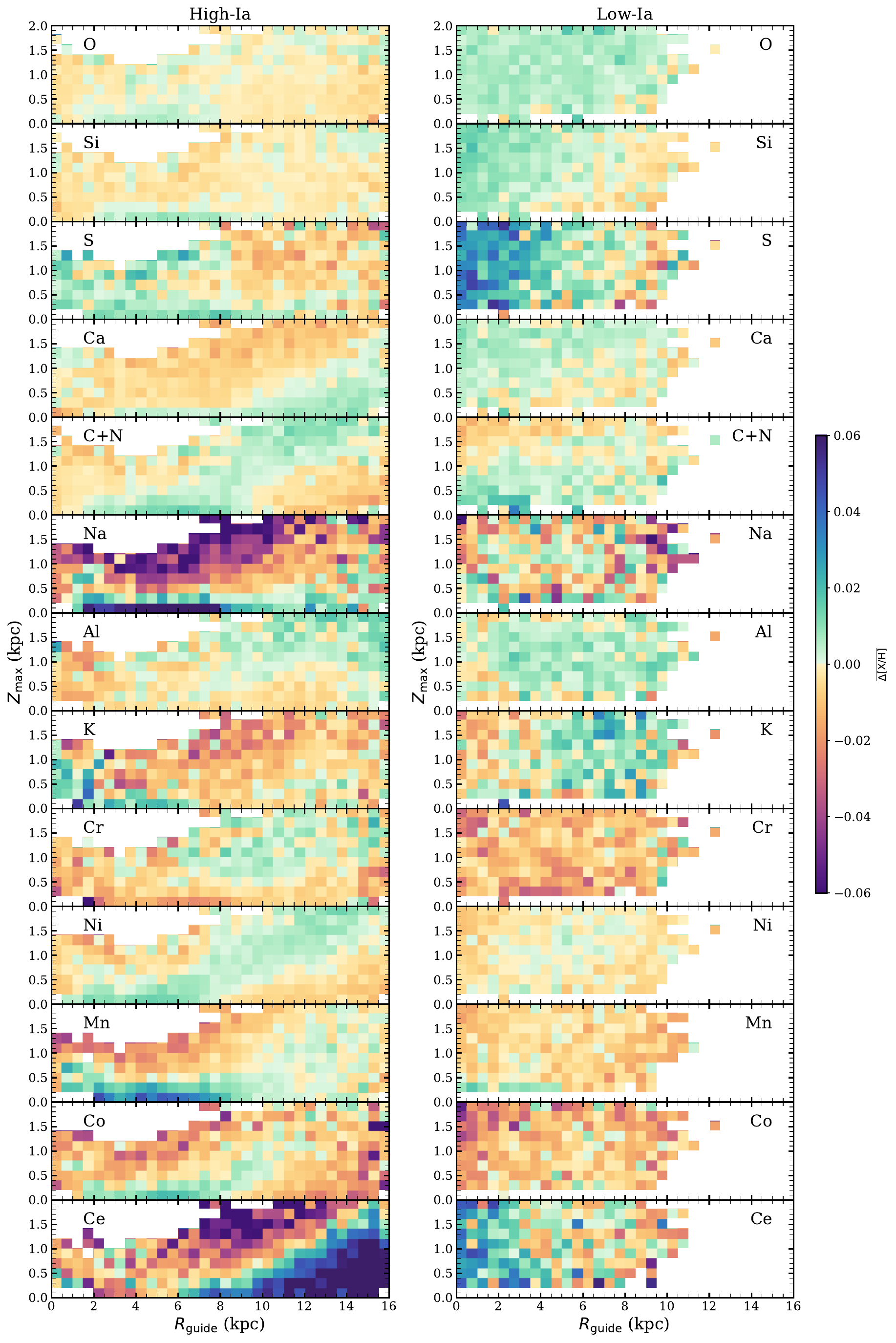}
    \caption{Maps of the average residual ($\Delta$[X/H] = [X/H]$_{\rm obs}$ - [X/H]$_{\rm pred}$) in $\Zm$ vs. $\Rg$ for 13 elements. We divide the population into high-Ia (left) and low-Ia (right). Positive average residuals are shown in pinks and oranges and negative average residuals are shown in green and blues. Bins with less than 20 stars have been excluded.}
    \label{fig:resid_map_Z_R}
\end{figure*}

After subtracting the KPM predictions from the observed abundances, we calculate average residuals ($\overline{\rm \Delta[X/H]}$) in bins of $\Rg$ and $\Zm$. In Figure~\ref{fig:resid_map_Z_R}, we plot the average residual values mapped over $\Rg$ from 0 to 16 kpc and $\Zm$ from 0 to 2 kpc at a resolution of 0.5 kpc in $\Rg$ and 0.2 kpc in $\Zm$ for the high-Ia (left) and low-Ia (right) populations. 
The average residual values are robust, as we typically average tens to hundreds of stars per bin. We do not include the residual Mg and Fe abundances, as they are zero for all $\Rg$ and $\Zm$ by construction. For other elements, a positive residual means that stars in this $\Rg$ and $\Zm$ bin have, on average a higher [X/H] than expected based on the trends on [X/H] with [Mg/H] and [Mg/Fe] present in the sample as a whole (Figure~\ref{fig:MgFe_map_Z_R}). A negative residual implies the opposite. Reference the top row of Figure~\ref{fig:density_maps.pdf} for the density distributions of stars on this grid.
Within each panel, the residual structure is visually obvious, both in elements like Na and Mn where the average residuals are large ($\sim0.06$ dex) and in elements like Si and Ca where the average residuals are small ($\sim0.02$ dex). For most elements, the peak magnitude of the average residual is near or below the abundance precision within APOGEE, highlighting the need for statistical analysis of a large dataset to search for residual correlations with orbital parameters. 

The patterns of the residual structure differ between elements and between the high-Ia and low-Ia stars, though some elements within each population display the same qualitative trends. Below we categorize like elements and describe their general residual structure. In the high-Ia population:
\begin{itemize}
    \item S and K have a positive to negative residual gradient with increasing $\Rg$ that is relatively uniform with $\Zm$. The transition from positive to negative occurs near 9 kpc in S and 3 kpc in K. Cr and Al show a similar trend of opposite sign to K at larger $\Zm$ and tend to have negative residuals at $\Zm < 0.5$.
    \item The Na and Mn residuals have a positive to negative gradient with increasing $\Zm$. This trend is the most obvious mid-plane for $\Rg \lesssim 8$. The transition from positive to negative residuals occurs near 0.4 kpc. While the $\Zm$ residual trend is relatively uniform with $\Rg$ for Na, the Mn residual is preferentially positive. The residuals for Si and O are small but show a similar spatial pattern. 
    \item Ca and Ce have a radial residual trend similar to S, but with the transition from positive to negative residuals occurring at larger $\Zm$ with increasing $\Rg$. This gives the appearance of diagonal stripes.
    \item In C+N, Ni, and Co a different diagonally residual structure arises. Here, stars at low $\Zm$ have a sinusoidal-like residual trend with $\Rg$ oscillating from negative to positive to negative again.  At $\Rg \approx 4$ kpc, the vertical residual trend is positive-to-negative, but at $\Rg \approx 10$ it is negative-to-positive.
\end{itemize}
In the low-Ia population:
\begin{itemize}
    \item The O, Ca, and Al residuals are positive in most regions. 
    \item The Si, S, and Ce residuals show a positive to negative radial gradient, from the inner to outer Galaxy, at all $\Zm$. K shows a similar but weaker trend in the opposite direction.
    \item C+N has a residual trend in $\Zm$ from positive to negative with increasing height. The transition in sign occurs at $\Zm\approx 1.5$.
    \item The residuals of the Fe peak elements, Cr, Ni, Mn, and Co, are negative in most regions. At low $\Zm$ the Mn residuals are positive and in Ni there is a faint, complex radial structure in the average resiudals.
    \item The residual structure of Na has no clear trend with $\Rg$ or $\Zm$.
\end{itemize}

For the most part, spatial structure is less well defined in the low-Ia residuals than the high-Ia residuals, with the radial residual gradient of S and Ce and the vertical residual gradient of C+N being the clearest. We note that the observed residual gradients in the low-Ia disk suggest that there may be a small radial metallicity gradients, in tension with previous works that find flat metallicity gradients in the low-Ia disk \citep[e.g.,][]{hayden2015, imig2023}

KPM predicts that the linear abundances (X/H) are linear functions of the linear abundances (Mg/H) and (Mg/Fe) (Equation~\ref{eq:kpm1}). The tendency of the low-Ia residuals to be weakly positive for $\alpha$-elements and weakly negative for Fe-peak elements indicates that the linear trends within the high-Ia population, which constitutes most of the sample, do not extrapolate perfectly to the (Mg/Fe) values of the low-Ia population. The high-Ia residual Ni, Mn, and Co maps resemblance of the average [Fe/Mg] in Figure~\ref{fig:MgFe_map_Z_R}, suggests that even within the high-Ia disk, the Fe abundances do not exactly track other Fe-peak elements.

\subsection{Residual Structure in Age vs. Guiding Radius}\label{subsec:Rvsage}

\begin{figure}
    \centering
    \includegraphics[width=1\linewidth]{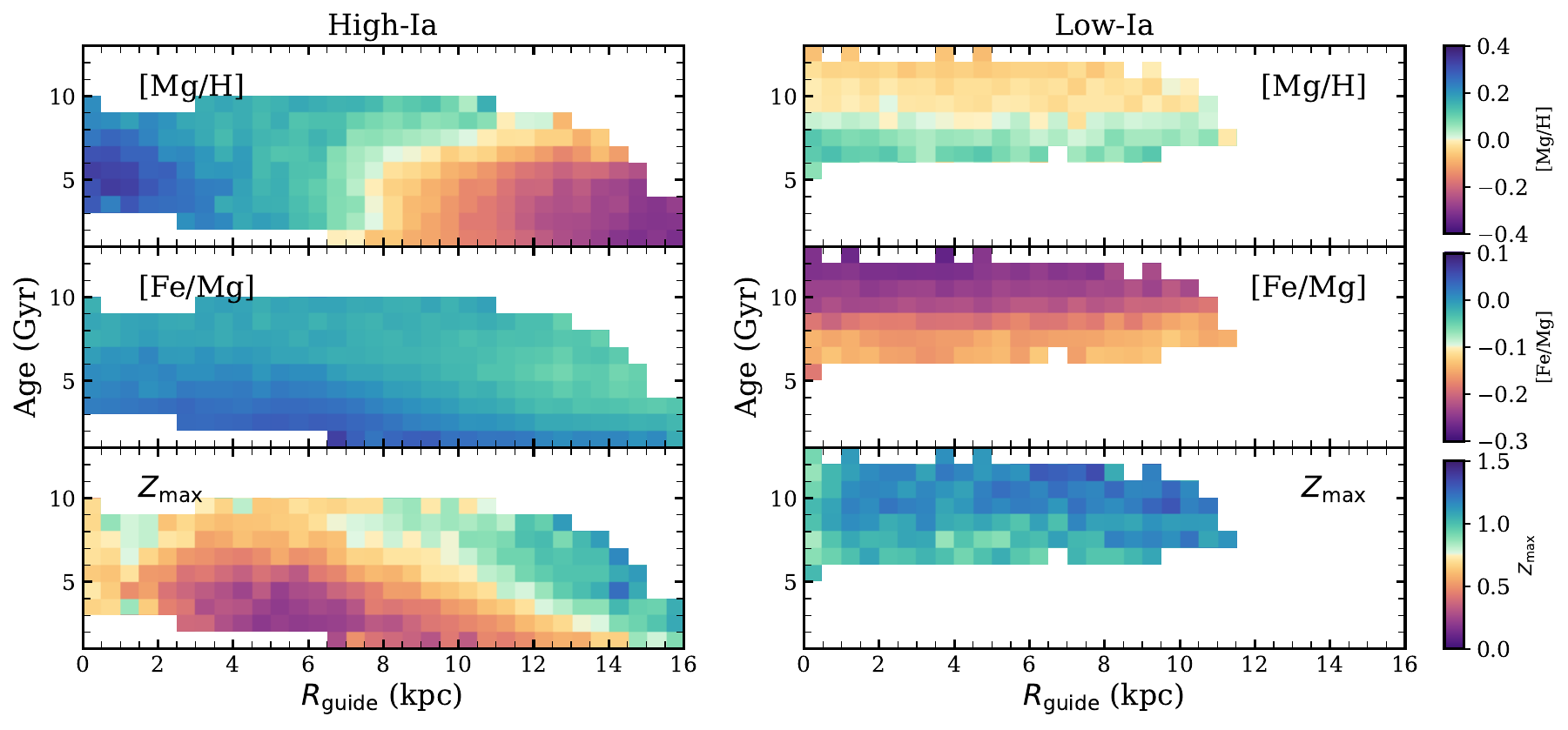}
    \caption{Maps of the average [Mg/H] (top), [Fe/Mg] (middle), and $\Zm$ (bottom) for the high-Ia (left) and low-Ia (right) populations in age vs. $\Rg$. Bins with less than 20 stars have been excluded.}
    \label{fig:MgFe_maps_age_R}
\end{figure}

In this Section, we turn to the residual structure in maps of age vs. $\Rg$. For reference, we plot the average [Mg/H], [Fe/Mg], and $\Zm$ in this space in Figure~\ref{fig:MgFe_maps_age_R}. Comparing with Figure~\ref{fig:MgFe_map_Z_R}, it is clear that some features in $\Zm$ map to age. Young stars (age $\lesssim 5$ Gyrs) tend to be in the midplane with $\Zm<0.5$ at $\Rg <8$ kpc, but they extend to larger $\Zm$ with increasing $\Rg$. 

\begin{figure*}
    \centering
    \includegraphics[width=1\linewidth]{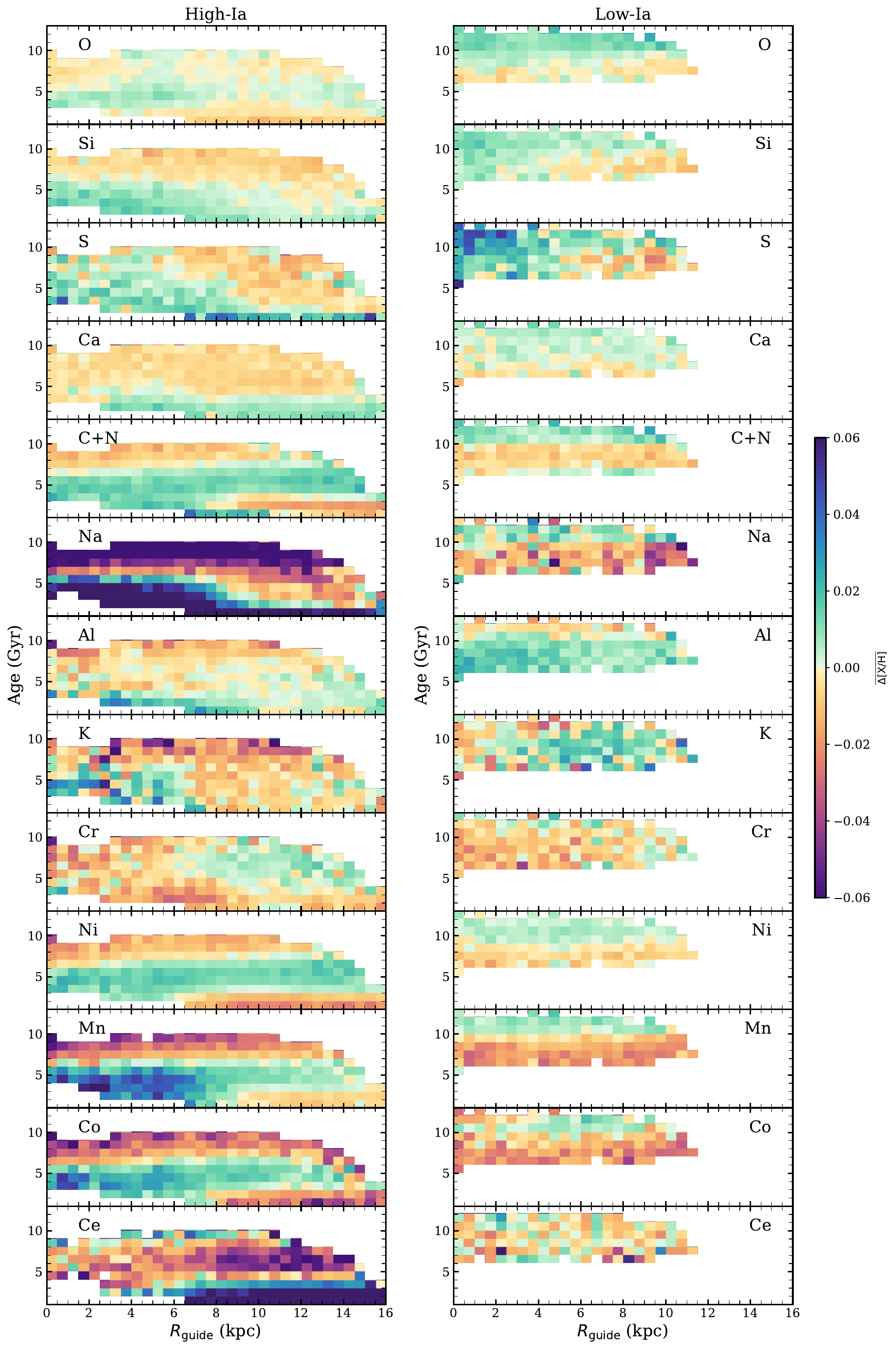}
    \caption{Same as Figure~\ref{fig:resid_map_Z_R} but for age vs. $\Rg$.}
    \label{fig:resid_map_age_R}
\end{figure*}

In Figure~\ref{fig:resid_map_age_R} we plot the average residual mapped over $\Rg$, as in Figure~\ref{fig:resid_map_Z_R}, but now with age from 1 to 13 Gyrs at a resolution of 1 Gyr on the y-axis. Reference the bottom row of Figure~\ref{fig:density_maps.pdf} for the density distributions of stars on this grid. 
Again, positive (negative) residuals imply that stars in a bin are on average higher (lower) in [X/H] than expected based on the overall sample trends.
As discussed in Section~\ref{sec:data}, \texttt{distmass} ages are only available for stars with $\logg>1$ and [Fe/H] $>-0.5$, about two-thirds of our sample. In age vs, $\Rg$ space, we also find clear structure with unique patterns. Broadly, the elements can be categorized by the observed residual structure in age vs. $\Rg$.
In the high-Ia population:
\begin{itemize}
    \item The Si, Ca, Na, Al, and Ce residuals have gradients in age that are relatively constant with $\Rg$. The residuals are positive at young ages and negative at old ages, with the transition occurring near 4 Gyrs for Ca, Al, and Ce and near 6 Gyrs for Si. The Na trend is a bit more complex, as a knee appears around 8 kpc. Here the residual zero-point shifts from 6 Gyrs to 3 Gyrs.
    \item The S and K residuals have radial gradients, with positive values in the inner Galaxy and negative values in the outer Galaxy. The transition from positive to negative occurs at increasing age with decreasing $\Rg$. Cr residuals show a similar structure with opposite sign.
    \item The residual structures in O, C+N, Ni, Mn, and Co are more complex. For these elements older stars (age $\gtrsim 7$ Gyrs) at all radii have negative residuals and younger stars at most radii have positive residuals--except for the youngest stars (age $\lesssim 4$ Gyrs) in the outer Galaxy ($\Rg \gtrsim 9$ kpc) where the residuals are negative. 
\end{itemize}
We observe that, for most elements, the fiducial model tends to overpredict the abundances of younger stars (age $\lesssim6$ Gyrs) and underpredict the abundances of older stars, especially at $\Rg \lesssim 8$ kpc.

In the low-Ia population:
\begin{itemize}
    \item O, Ca, C+N, Na, Al, Ni, Mn, and Co residuals show gradients with age of different signs and slopes.
    \item The Cr residuals are dominantly negative except for at the largest radii.
    \item The K residuals are dominantly positive except for at small radii.
    \item The S residuals show a clear radial gradient from positive in the inner Galaxy to negative at larger radii. Na has a similar trend for stars with ages of $6-10$ Gyrs, such that stars at larger radii have more negative residuals than stars in the inner Galaxy 
    \item The Ce residuals have no clear trends and appear random.
\end{itemize}
No element shows the same residual structure in its low-Ia population as in its high-Ia population, though many elements, such as Ca, C+N, Ni, and Mn show similar structures of opposite signs.

\begin{figure*}
    \centering
    \includegraphics[width=1\linewidth]{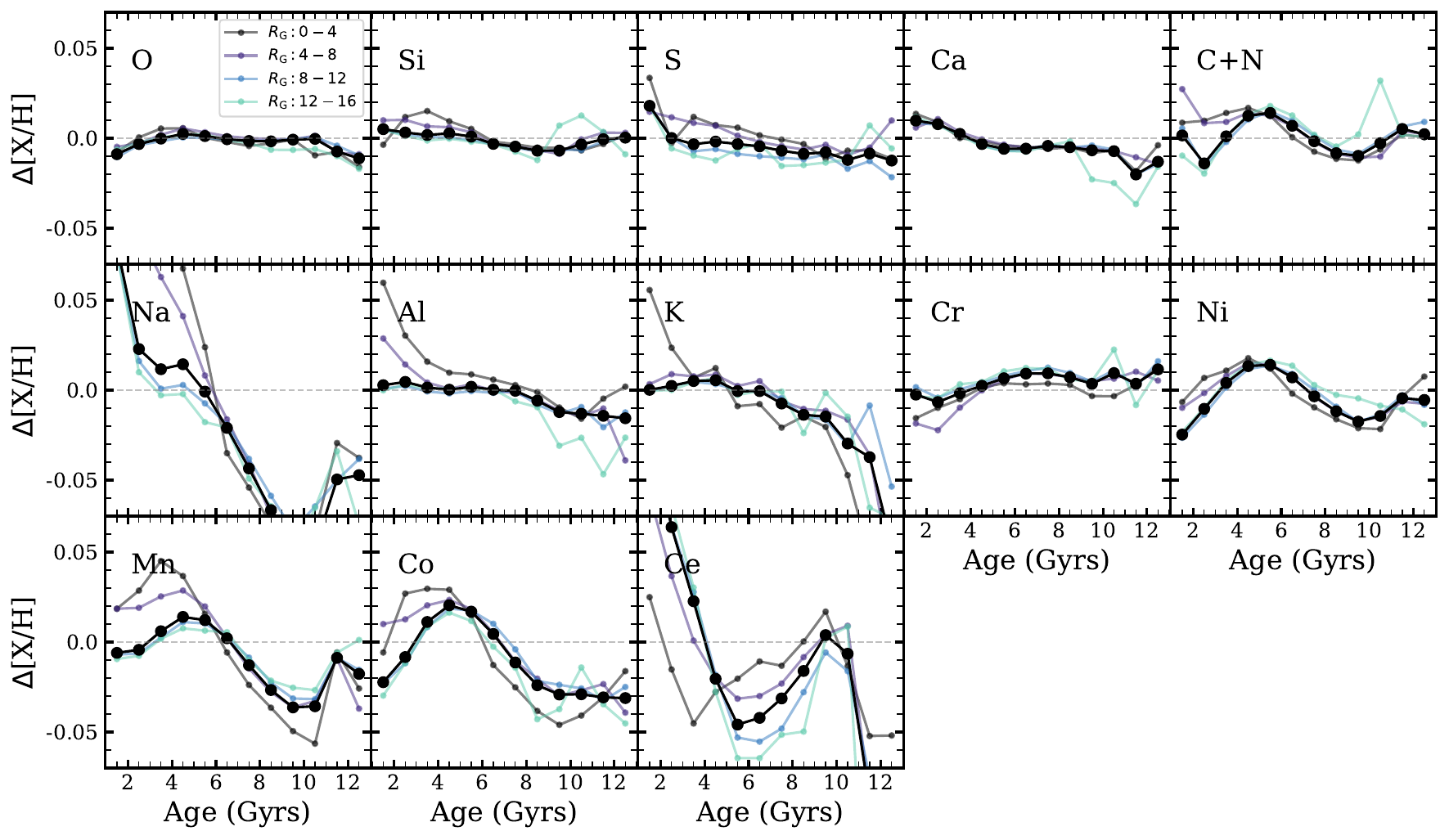}
    \caption{The age vs. average residual ($\Delta$[X/H]) for 13 abundance residuals in the full high-Ia population (black) and subsamples binned by increasing radius (dark blue to teal). Average residuals are calculated in 1 Gyr width bins from 1 to 13 Gyrs.}
    \label{fig:age_resid}
\end{figure*}

Of the three dimensions shown in Figures~\ref{fig:resid_map_Z_R} and~\ref{fig:resid_map_age_R}, the abundance residuals have the strongest 1D trends with age. In Figure~\ref{fig:age_resid}, we highlight the age vs. $\Delta[\rm X/H]$ relationship for the full high-Ia population and subsamples binned by $\Rg$. Many elements, including C+N, Ni, Mn, and Co, show sinusoidal-like curves at all radii with an amplitude of 0.02 dex or greater. Na and Ce show a strong linear rise in residuals at young ages. The slopes of the Na, Al, K, Mn, and Co trends are greatest in the inner Galactic regions. 
Previous two-parameter model analyses have also identified residual abundance trends with age. Of note, \citet{weinberg2022} find age trends with Na and Ce residuals, and S24 identify positive residuals of S, C+N, Na, and Ce in the youngest open clusters. We note that Ce, in particular, is produced through the slow neutron capture process ($s$-process; \citealp{arlandini1999}), and has been used as a chemical clock to estimate stellar ages \citep[e.g.,][]{ratcliffe2024} along with Y and Ba \citep[e.g.,][]{horta2022, hayden2022}. 

\subsection{The Dimensionality of the Residual Structure in Age and Radius} \label{subsec:PCA}

As discussed throughout this paper, Mg and Fe are strong predictors of stellar abundances to a precision of roughly 0.05 dex. While these two components dominate, it is clear that additional structure lies within the residuals, suggesting that, at small scales, the APOGEE abundances have a dimensionality greater than two. Recent PCA, clustering, and latent space analysis of APOGEE and GALAH data support this, as works find that 5 to 10 eigenvectors are needed to fully explain the observed abundances \citep{ting2012, price2018, casey2019, ratcliffe2020, ratcliffe2022}, highlighting the potential impact of SFH on the detailed abundance structure observed in the MW.

\begin{figure}
    \centering
    \includegraphics[width=\linewidth]{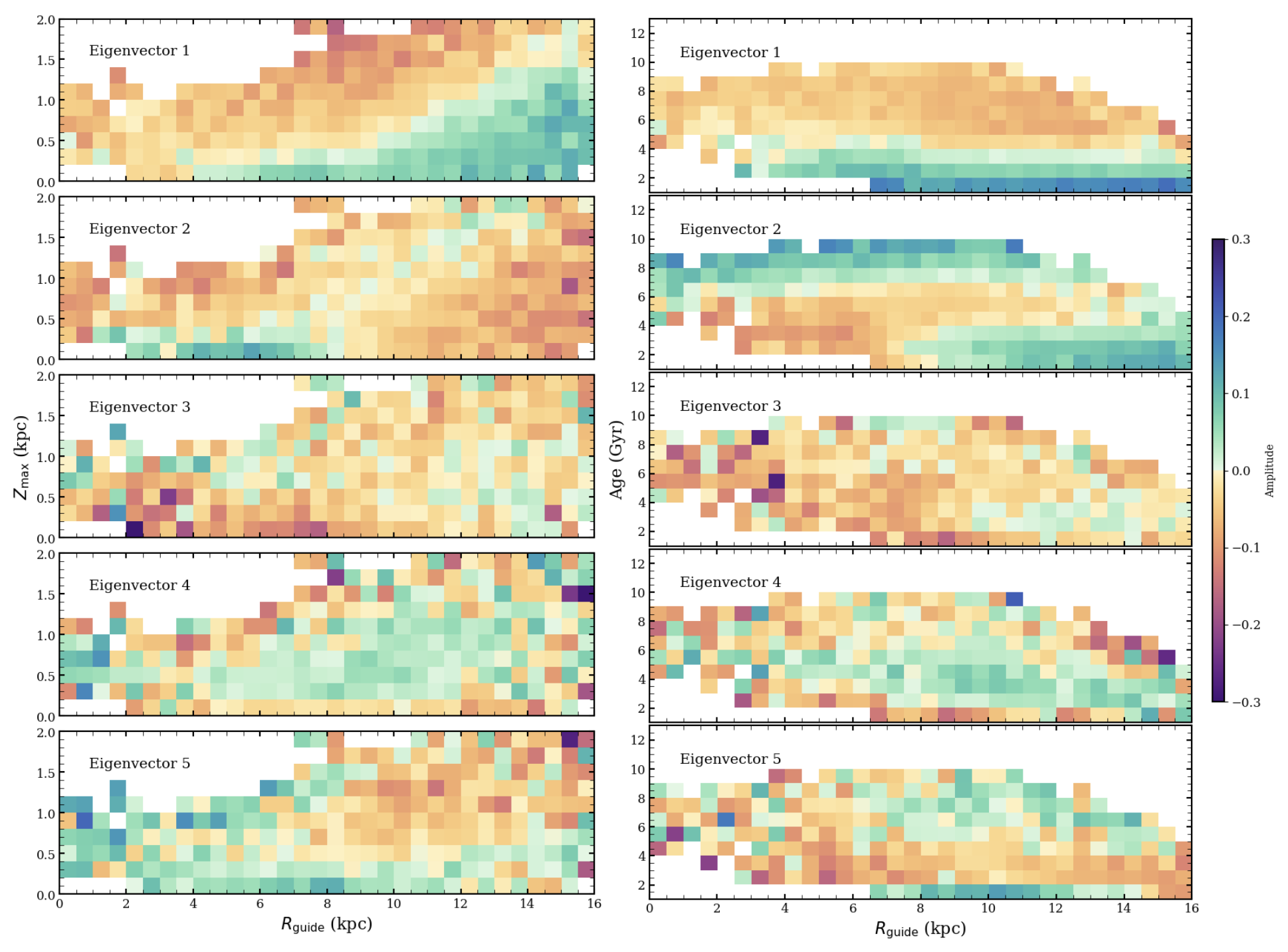}
    \caption{Eigenvectors of a PCA fit to the residual abundances of the high-Ia population in $\Zm$ vs. $\Rg$ (left) and age vs. $\Rg$ (right). The amplitude and sign of the component colors each cell. Note that the color bars differ in each panel.}
    \label{fig:PCA}
\end{figure}

To quantify the dimensionality of the residual structure after abundance correlations with Mg and Fe have been removed, we conduct a PCA decomposition of the high-Ia residuals in $\Zm$ vs. $\Rg$ and age vs. $\Rg$ space, separately. We choose to work only with the high-Ia stars as they show the strongest and most featured residual trends.
PCA---principal components analysis---in this context seeks to find a low rank representation of the element residual maps such that each map can be seen as a linear combination of a few eigenmaps (eigenvectors hereafter).
The eigenvectors are chosen to be orthogonal, and ordered by how much of the variance in the original data they can explain. 
We use the PCA implementation from scikit-learn \citep{scikit-learn}, simultaneously fitting the binned average residual abundance ``images'' for 13 elements (excluding Mg and Fe) on the same grids as shown in Figures~\ref{fig:resid_map_Z_R} and~\ref{fig:resid_map_age_R}.
We show the first five eigenvectors of the PCAs, capable of describing 95\% of the total abundance residual variance, in Figure~\ref{fig:PCA}. The first three components can describe 90\% of the total variance. 
Note that the PCA decomposition here is different from the dimensionality estimates cited above, which seek to characterize the effective number of dimensions in abundance space itself. Here we ask how many dimensions in $\Zm$ vs. $\Rg$ and age vs. $\Rg$ are required to explain the residual abundance patterns across many elements.

In the eigenvectors, we see clear structure that resembles the residual maps from Figures~\ref{fig:resid_map_Z_R} and~\ref{fig:resid_map_age_R}. In $\Zm$ vs. $\Rg$, the first eigenvector shows a strong gradient with $\Zm$ in the outer galaxy with the transition from positive to negative occurring at larger $\Zm$ for larger $\Rg$. The second eigenvector has a stronger positive to negative gradient in $\Zm$ in the inner Galaxy, and an oscillating residual trend in the outer Galaxy. The third eigenvector is dominated by a weak linear radial trend that is strongest in the inner galaxy. The fourth eigenvector shows a uniform value across most of the disk, with some complex structure around the edge of the image. Finally, the fifth eigenvector shows a radial trend with a sign transition near 8 kpc.

The eigenvectors in age vs. $\Rg$ partly mirror those in $\Zm$ vs. $\Rg$, with a translation of average height to average ages. However, in age vs. $\Rg$ the first eigenvector primarily shows a linear gradient in age, with the sign transition at 3 Gyr and little trend in $\Rg$. The second eigenvector has a more complex spatial structure, showing an oscillating amplitude trend in age, with sign transitions near 4 and 8 Gyrs. A radial trend from negative to positive amplitudes is also apparent at $4 < \Rg < 8$ kpc for young stars (age $< 6$ Gyrs). The third eigenvector is dominantly a weak radial gradient, with more negative amplitudes in the inner Galaxy and smaller amplitudes in the outer Galaxy. Eigenvector four is uniformly positive across most ages and radii, with negative features around the image edge. The final eigenvector is more complex, with radial and age trends that are not uniform across the parameter space.

\begin{figure}
    \centering
    \includegraphics[width=0.8\linewidth]{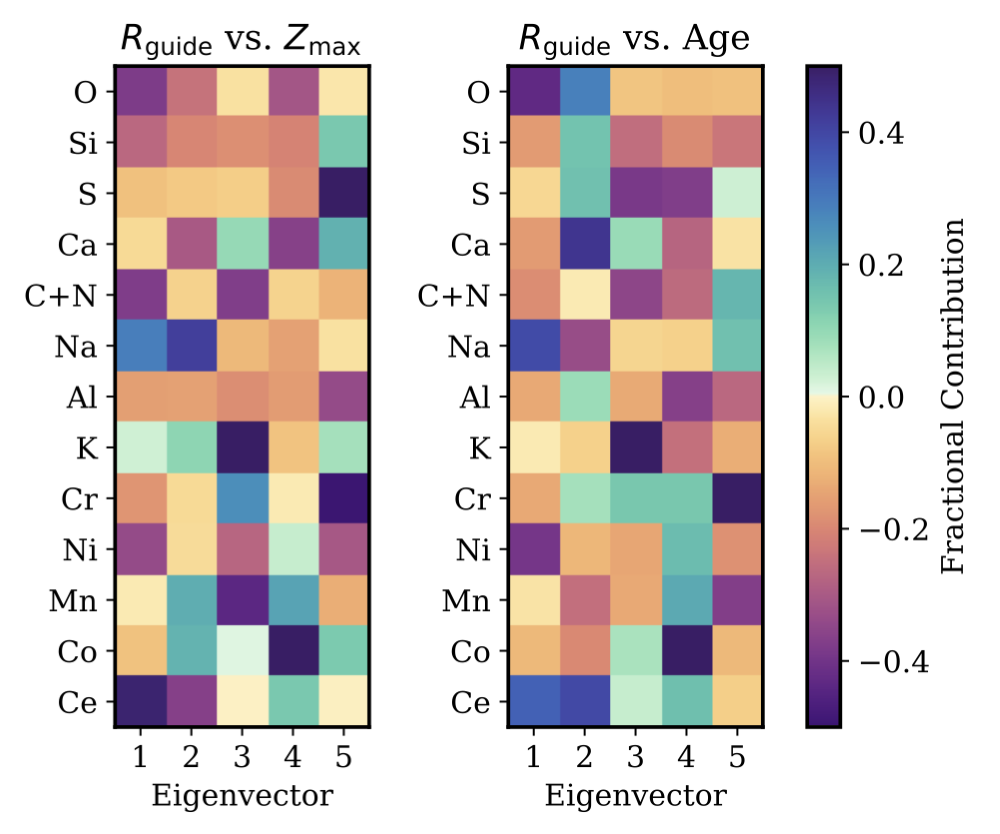}
    \caption{Fractional contribution, including sign, of each eigenvector to each element for the PCA decomposition in $\Zm$ vs. $\Rg$ (left) and the PCA decomposition in age vs. $\Rg$ (right).}
    \label{fig:PCA_comp}
\end{figure}

Each element's abundance residual structure can be nearly recreated by summing a subset of the weighted components. In Figure~\ref{fig:PCA_comp}, we show the fractional contribution of each component to a given element's residual structure, colored by weight and sign. Broadly, in $\Zm$ vs. $\Rg$, we find that the first component strongly contributes to O, Si, C+N, Na, Ni, and Ce; the second to Ca, Na, and Ce; the third to C+N, K, and Mn; the fourth to Ca and Co; and the fifth to S, Al, and Cr. In age vs. $\Rg$, we find that the first component strongly contributes to O, Na, Ni, and Ce; the second to Ca, Na, and Ce; the third to S, C+N, and K; the fourth to S, Al, and Co; and the fifth to Cr and Mn. There are notable similarities between the PCA weights in both parameter spaces. We note that the element groups seen here do not strongly resemble the qualitative groups described in Sections~\ref{subsec:RvsZ} or ~\ref{subsec:Rvsage}. This highlights that while the residual structure may look alike between sets of elements, they differ in details such as the specific location of the residual zero-points and the residual gradient slopes. Components that contribute substantially to only one element suggest that these elements (K, Co, Cr) are not like the others and may have some unique enrichment process or residuals that are dominated by unique systematic errors.

Overall, the PCA decomposition highlights the clear residual structure that correlates with stellar age, height, and radius. The residual variance for 13 elements can be captured by just three to five eigenvectors, showing that the same underlying residual structure exists in many elements. This supports the conclusions of previous abundance decomposition analyses that the Galaxy's stellar abundances have more than two dimensions.

\section{Potential Causes of Abundance Residual Structure} \label{sec:causes_resids}

A striking finding from APOGEE's ``chemical cartography'' of the Milky Way is that median abundance trends ([Mg/Fe] vs. [Fe/H] and [X/Mg] vs. [Mg/H]) are nearly universal throughout the disk and bulge, provided one separates the low-Ia and high-Ia populations \citep{hayden2015,weinberg2019,griffith2021a}.
This universality across zones with radically different star formation and gas accretion histories implies that the median trends are governed by stellar nucleosynthesis rather than galactic scale physics, motivating the KPM model.  
The most striking general result in Figures~\ref{fig:resid_map_Z_R} and~\ref{fig:resid_map_age_R} is that the {\it residuals} from the $K=2$ model are coherent in Galactic location and age, a clear departure from the universal behavior of the median abundance
trends.  
The magnitude and spatial coherence of these average residuals is much larger than would be expected if each star was drawn independently from an underlying stochastic distribution.  Thus, it appears that the deviations from KPM are coupled to the distinct histories of different Galactic zones, in contrast to the median trends themselves.

Within the Milky Way disk, stars span $\sim 1$ dex in [Mg/H] and $\sim 0.3$ dex in [Mg/Fe]. The study of average residuals allows us to map the much subtler, $\sim 0.02-0.1$ dex variations of abundance ratios that remain after we remove the trends tied to [Mg/H] and [Mg/Fe].  While there have been 
numerous theoretical studies of the Galaxy's chemo-dynamical structure in metallicity and [Mg/Fe], there is little existing theory for interpreting the lower amplitude structure in abundance ratios measured here.  Many factors could produce deviations from the simplest KPM predictions, including observational systematics, variations in IMF-averaged yields, and variations in the relative contribution of additional enrichment channels beyond CCSN
and SNIa. For elements with metallicity-dependent yields, abundance variations can arise because different regions may take different paths to the same [Fe/Mg] and [Mg/H] (e.g., \citealt{johnson2020,johnson2023,spitoni2024}).

In the following subsections, we revisit the influence of abundance systematics and additional processes on the residual abundances. We take a deeper dive into the impact of dilution, as it is easily integrated into KPM and allows for a new type of model flexibility. In future work, we hope to explore the impact of the KPM assumptions, such as linearity and [Mg/H] metallicity dependence in more depth.

\subsection{Systematic Trends with Stellar Parameters}\label{subsec:systematics}

We have identified structure in the residual abundances away from the fiducial KPM model that correlates with $\Rg$, $\Zm$, and age. This structure could be caused by physical differences in the abundances of stars of different ages or located on different Galactic orbits, which cannot be explained by a $K=2$ process model. However, the structure could also be caused by abundance correlations with nuisance parameters, such as $\logg$, $\teff$, and radial velocity (RV) that have eluded our calibration, if these parameters also correlate with orbits and/or ages. 

In Section~\ref{subsec:calibration}, we discuss the known abundance systematics that correlate with $\logg$ and efforts by S24 to implement a calibration scheme that removes these effects. Because of this calibration, we do not expect $\logg$ or $\teff$ systematics to be the dominant sources of residual structure in Figures~\ref{fig:resid_map_Z_R} and ~\ref{fig:resid_map_age_R}. We do not apply corrections for systematic abundance trends with RV, which might arise due to weak diffuse interstellar bands that contaminate absorption features for stars of specific RV along specific lines of sight \citep{ness2022, mickinnon2024}. 

We calculate the correlation ($r$) between the residual [X/H] abundances and nuisance stellar parameters to check for systematic abundance trends with $\logg$, $\teff$, and RV. Co, Mn, and Ce are the only elements with $|r|>0.1$ for one or more stellar parameters. Mn, one of the elements with a large $\logg$ calibration in S24, has the most significant remaining correlations with the nuisance parameters ($|r|\approx 0.2$), with the coolest stars systematically under-predicted by KPM. K and Na display strong bands of stars with larger residuals at RV near 100 km/s. Most element-parameter pairs, however, have small correlations ($|r|<0.03$). 

\begin{figure*}
    \centering
    \includegraphics[width=1\linewidth]{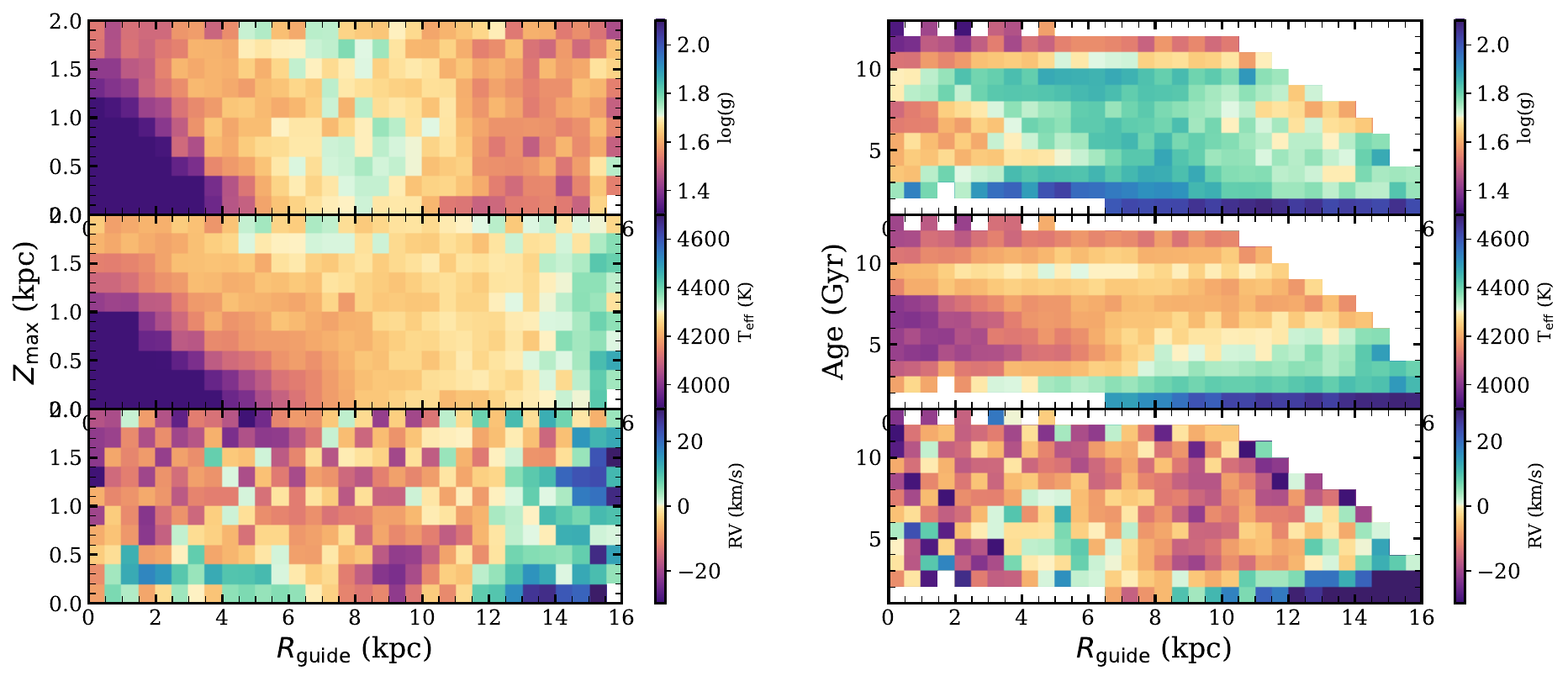}
    \caption{Maps of the average $\logg$ (top), $\teff$ (middle), and RV (bottom) in $\Zm$ vs. $\Rg$ (left) and age vs. $\Rg$ (right). Bins with less than 20 stars have been excluded.}
    \label{fig:nuis_map}
\end{figure*}

To further explore the potential impacts of systematic abundance trends with stellar parameters, we recreate the maps from Figure~\ref{fig:resid_map_Z_R} and Figure~\ref{fig:resid_map_age_R}, now colored by average $\logg$, $\teff$, and RV in Figure~\ref{fig:nuis_map}. We combine the high-Ia and low-Ia populations, as the average stellar parameters are consistent in overlapping $\Zm$ vs. $\Rg$ and age vs. $\Rg$ regions. If the average residual structure resembles the average stellar parameter structure in these coordinates, the residual structure may be biased.

In $\Zm$ vs. $\Rg$, the average $\logg$ and $\teff$ tend to be lower in the inner Galaxy at low $\Zm$, and increase with increasing radius and height. This pattern slightly resembles that of Al and K in the high-Ia population. The average RV appears random for $\Rg<12$ and $\Zm<0.5$, and tends to be positive and larger magnitude at low $\Zm$ and large $\Rg$. This pattern does not obviously resemble any residual structure in Figure~\ref{fig:resid_map_Z_R}.

In age vs. $\Rg$, the average $\logg$, $\teff$, and RV tend to be larger at young ages (age$< 3$ Gyrs). This is expected, as the population of evolved stars moves down the HR diagram to lower luminosities and temperatures with increasing age. At older ages, $\logg$ and $\teff$ show some banding and the RV appears random. In the high-Ia panels of Figure~\ref{fig:resid_map_age_R}, we see many elements with bands of positive (Na, Al, Ce) and negative (C+N, Ni, Mn, Co) residuals at young ages in the outer disk that resemble the stellar parameter feature. If the abundance residuals for these elements are correlated with $\logg$, $\teff$, and/or RV, the young star abundance residuals may be skewed by systematic effects. Of these elements, Al, Ni, Mn, Co, and Ce have slight correlations with $\teff$ ($|r|>0.05$), though this is most visually apparent in cool stars. Systematic abundance residual trends with nuisance stellar parameters may contribute to the residual structure for these elements, but they are not obviously the dominant cause of the structure. We proceed in our analysis without applying additional abundance calibrations.

\subsection{An AGB Processes} \label{subsec:agb}

In the fiducial $K=2$ model, the first process is fixed to Mg, to capture CCSN enrichment, and the second to Fe, to capture SNIa enrichment. While this simple nucleosynthetic model is effective, it excludes additional known enrichment channels such as AGB stars \citep[e.g.,][]{simmerer2004, karakas2016} and subclasses of SNIa \citep[e.g.,][]{reyes2020, gronow2021}. As explored in depth in G24, KPM can be extended to include as many processes as the user specifies, so long as each process is fixed to an element. In G24, adding two additional processes, fixed to Ce and Mn, improved the statistical ability of the model to fit the data, but the additional processes were not clearly representative of AGB stars or a distinct class of SNIa. As no abundance calibrations were included in their analysis, they conjectured that the additional model flexibility may have been fitting systematics rather than physical features. 

Here, we revisit a simplified implementation of a $K=3$ model with $\logg$ calibrated abundances. We fix the third process to Ce, an element produced in AGB stars, such that
\begin{equation}\label{eq:q_3}
    q_{{\rm 3, Mg}}^{\,Z} = 0, \quad q_{3,{\rm Fe}}^{\,Z} = 0, \quad q_{3,{\rm Ce}}^{\,Z} = 1,
\end{equation}
along with constraints from Equations~\ref{eq:qcc_mg_solar} and~\ref{eq:qcc_fe_solar}. As in the fiducial model, process vector components are fit to all elements and process amplitudes are fit to a subset of elements for all stars---now Mg, Fe, and Ce. We initialize the new model at the best-fit parameters of the fiducial model and slightly regularize the third process to achieve the best results.

In this simplified implementation of a $K=3$ model, we find that the additional process only significantly improves the residuals of Ce. This suggests that no other elements have Ce-like enrichment that is not captured by Mg and Fe. Consequently, we do not see any changes to the residual maps in $\Zm$ vs. $\Rg$ or age vs. $\Rg$ for elements other than Ce. From this test, we conclude that AGB stars are not responsible for the residual structure observed in elements other than Ce. This does not mean that AGB stars do not contribute to the enrichment of these elements, but variation in AGB enrichment at a given [Mg/H] and [Fe/Mg] does not appear to drive the observed residual abundance structure. We repeat this analysis with other elements, including C+N and Mn, finding the same results.

\subsection{A Massive Dilution Event} \label{subsec:dilution}

During a dilution event, pristine gas is added to the galaxy, increasing the amount of H while leaving the [X/Mg] abundances unchanged. This pushes back the metallicity to a lower value for stars that would have otherwise been born at higher metallicity if no inflow had occurred. Many recent comparisons of galactic chemical evolution models to Galactic abundances have concluded that a mass inflow (perhaps from the GSE merger e.g., \citealp{nissen2010, helmi2018, hayes2018}), is at least in part responsible for creating the high-Ia sequence \citep[e.g.,][]{chiappini1997, spitoni2019, spitoni2024}. The MW likely formed the low-Ia population, an evolutionary sequence, first before a large inflow diluted the super-solar metallicity gas back to a lower metallicity. Observed radial abundance and age gradients suggest that this dilution was strongest in the outer Galaxy \citep[e.g.,][]{mackereth2017}. Beyond major merger models, additional works have explored the potential for multiple dilution events with three-infall models \citep{micali2018, spitoni2022}.

Under the fiducial KPM model, we assume that each star is well described by the process vector components at its current [Mg/H], but if that star was born after a dilution event, it may be better fit by process vector components associated with a higher [Mg/H]. To capture this, we introduce a global dilution parameter, $\Delta$, which reduces (or increases) the [X/H] abundance while allowing the process vector components to reference a higher (or lower) pre-dilution metallicity, such that Equation~\ref{eq:kpm1} becomes
\begin{equation}\label{eq:kpm_delta}
     m_{ij} = \log_{10} \biggl( \sum_{n=1}^K A_{i}^n \, q_{n,j}(Z + \Delta) \biggr) - \Delta.
\end{equation}
The dilution parameter has no element dependence and impacts all abundance ratios uniformly. The $\Delta$ value is fixed by the user when initializing the model. Here, we allow $\Delta$ to be positive or negative as the fiducial model represents ``average'' enrichment. Thus, a negative $\Delta$ implies a star needs less dilution than the fiducial model while a positive $\Delta$ suggests a star needs more dilution than the fiducial model.

To fit a dilution coefficient to every star, we re-calculate the best-fit process amplitudes on a grid of potential $\Delta$ values from $-0.3$ to $0.3$ dex while fixing the process vector components to those found from the fiducial model. We calculate the $\chi^2$ value for every model and adopt the $\Delta$ value at the minimum $\chi^2$ as the best fit value. We expect the addition of the dilution parameter to most strongly impact elements with metallicity-dependent process vector components, such as Na and Mn, where a small change in reference metallically could more significantly change the KPM predicted abundance. We calculate the best $\Delta$ values for all stars, but focus on the high-Ia population.

\begin{figure}
    \centering
    \includegraphics[width=1\linewidth]{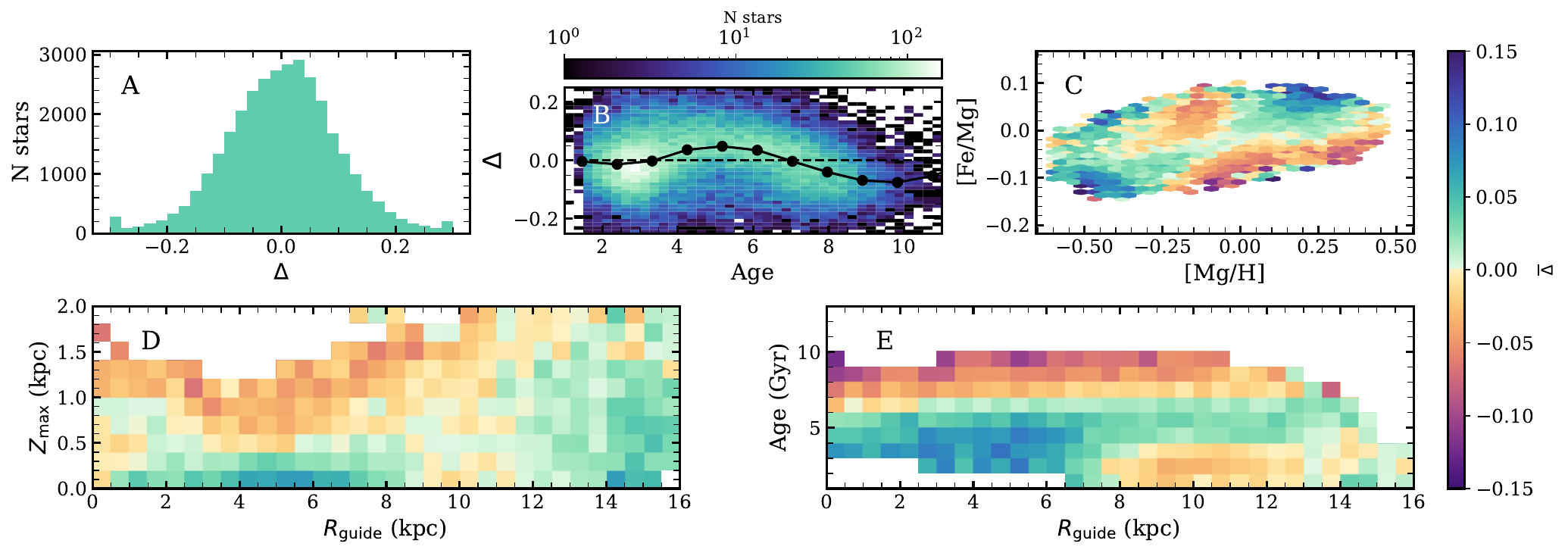}
    \caption{Plots of the best-fit dilution coefficient for the high-Ia population. A: Distribution of best fit $\Delta$ values. B: Distribution of $\Delta$ with stellar age. C: Average $\Delta$ in [Fe/Mg] vs. [Mg/H]. D: Average $\Delta$ in $\Zm$ vs. $\Rg$ E: Average $\Delta$ in age vs. $\Rg$. In panels D and E, bins with less than 20 stars have been excluded.}
    \label{fig:deltas}
\end{figure}

In Figure~\ref{fig:deltas}, we plot the best fit $\Delta$ values for the high-Ia population in many dimensions. Overall, we find that a range of positive and negative $\Delta$s are preferred, with a slight preference for positive values (panel A). A star's best fit $\Delta$ is correlated with its C+N residuals and age, such that the oldest stars prefer negative $\Delta$ and stars near 5 Gyrs prefer positive $\Delta$ (panel B). The dilution parameter has a complex relationship with metallicity, but overall we find that the most metal-poor high-Ia stars prefer positive $\Delta$ along with stars at super-solar [Mg/H] and [Fe/Mg] (panel C). In age vs. $\Rg$, we again see that the older stars prefer negative $\Delta$ while the younger stars ($<6$ Gyrs) prefer positive $\Delta$, except for the youngest stars from $\Rg$ of 8 to 14 kpc (panel E). Panels E of Figure~\ref{fig:deltas} strongly resemble the second PCA eigenvector from Figure~\ref{fig:PCA}, suggesting that the dilution parameter may be capturing a fundamental piece of the residual abundance structure in age vs. $\Rg$.

\begin{figure}
    \centering
    \includegraphics[width=1\linewidth]{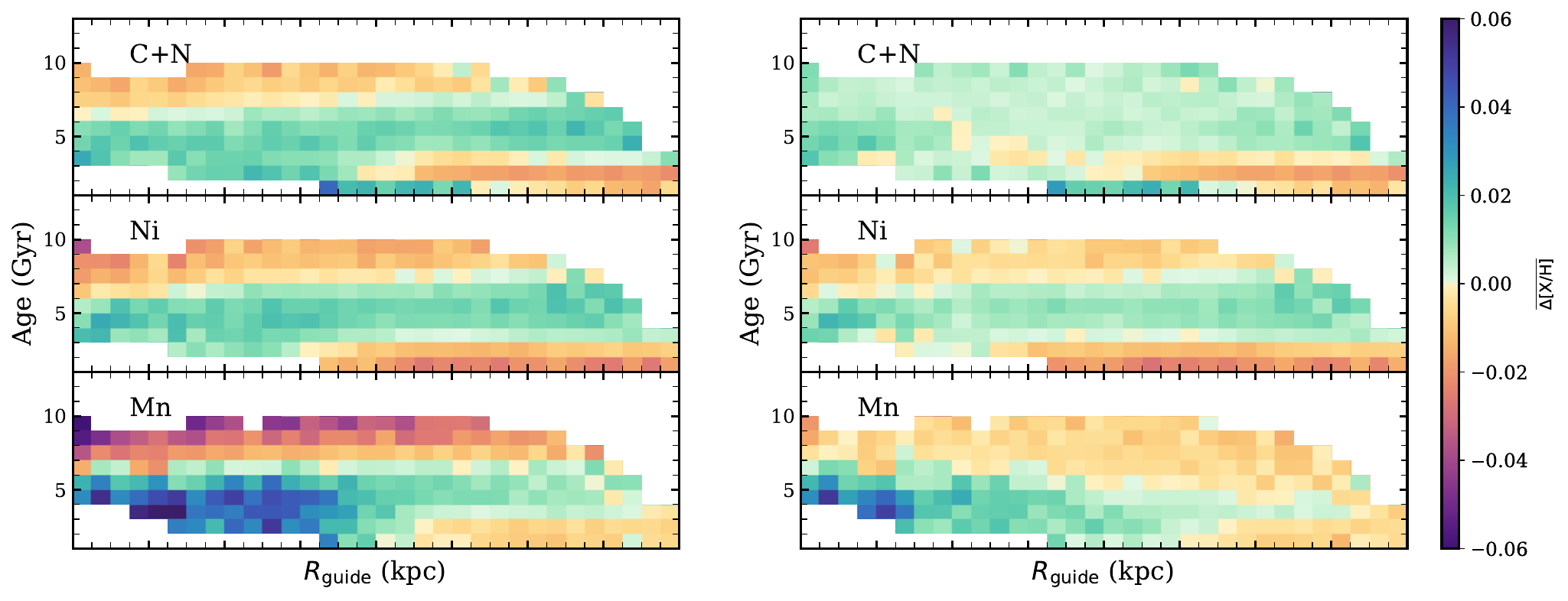}
    \caption{Average residual maps of the fiducial model (left) and dilution model (right) for the high-Ia population in age vs. $\Rg$.}
    \label{fig:resid_map_age_R_diff}
\end{figure}

The additional dimension of dilution statistically improves the KPM fits, with the greatest impact on C+N, Na, Ni, and Mn. We find that dilution captures some of the observed residual structure, but does not fully explain it. This is shown in Fig~\ref{fig:resid_map_age_R_diff}, where the amplitude of the average high-Ia C+N, Ni, and Mn residuals in age vs. $\Rg$ are larger for the fiducial model than the dilution model. The reduction in the amplitudes of the residuals for certain elements but not for others is likely tied to the metallicity dependence of the integrated yields. We expect the amplitude of the C+N, Na, Ni, and Mn residuals to be reduced if their metallicity dependence in delayed enrichment channels, such as SNIa (or AGB stars for Na, see \citealp{griffith2019}), is very high. Elements whose residuals are not improved by the addition of dilution must then have weaker metallicity dependence in the same delayed processes. 

In a realistic Galactic model, we would expect dilution to be similar for all stars born at a similar time and Galactic location. 
Fitting $\Delta$ parameters for individual stars and taking their average gives guidance to the kind of dilution event that might explain the observed residual structure. Figure~\ref{fig:deltas} suggests a dilution event approximately 5 Gyr ago, strongest in the 
inner Galaxy, and producing about 0.05-0.1 dex reduction in ISM abundances. We leave the task of constructing and testing models of such an event to future work.

\subsection{The Assumptions of the KPM Model}

In addition to the model assumptions about the number of nucleosynthetic processes and inflows discussed in detail in the preceding sections, other assumptions built into KPM could influence the observed residual structure. While exploring the implications of these model limitations is outside the scope of this paper, we enumerate them here. 

\textbf{1. [Mg/H] yield dependence:} In the fiducial implementation of KPM, the process vector components for all elements but Mg and Fe are allowed to vary with metallicity, where [Mg/H] is the assumed metallicity parameter. In practice, some elements may have yield dependencies on other [X/H] abundances, leading to a metallicity dependence that cannot be fully captured in the fiducial model. This is preliminarily explored with a similar two-parameter model fit to the local dwarf satellites in \citet{hasselquist2024}. They propose that elements formed during carbon and neon burning (e.g., Al and Na \citealp{truran1971}) may be dependent on [C+N/H] rather than [Mg/H]. They find some evidence of non-[Mg/H] metallicity dependence in the two-process residuals for Sagittarius and the LMC. Based on their results, we investigate the correlation between $\Delta$[X/H] and [C+N/H]. If the abundances of some element X have a metallicity dependence on C+N instead of Mg, then we would expect its residuals to correlate with the preferred reference elements. We find evidence of residual correlation with [C+N/H] and the Si, Na, Cr, Ni, Mn, Co, and Ce residuals, and find that these correlations are strongest at super-solar [Mg/H]. We hope to expand KPM to include more flexible metallicity dependence when fitting the process vector components in future work.

\begin{figure}
    \centering
    \includegraphics[width=1\linewidth]{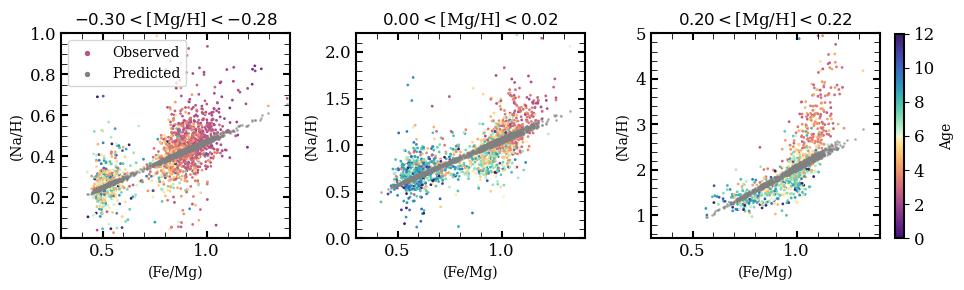}
    \caption{Linear (Na/H) vs. (Fe/H) for three bins of [Mg/H]. In each panel, stars are colored by their age. We plot the KPM predicted abundances in grey to show the model linearity in small bins of [Mg/H].}
    \label{fig:linear-Na}
\end{figure}

\textbf{2. Linearity:} A foundational assumption of KPM is that the linear (X/H) abundances of a star can be described as the linear combination of $K$ processes (Equation~\ref{eq:kpm1}). This assumption is likely incorrect in detail for some elements. To demonstrate, in Figure~\ref{fig:linear-Na} we plot the observed (Na/H) vs. (Fe/Mg) in three small ($0.02$ dex) bins of [Mg/H], $-0.3$ to $-0.28$, $0.0$ to $0.02$, and $0.2$ to $0.22$, colored by age. By fitting for the two amplitudes ($A_1$ and $A_2$), we can in principle fill any two dimensional plane of abundances with data. However, here we isolate a small range of [Mg/H], effectively conditioning on one abundance. Our predicted Na abundances must then be linear in this space (predictions shown in grey). If the model assumption is correct, the observed stars should follow the same linear trend, with some scatter from observational and systematic abundance errors. While the observed abundances are roughly linear in the low-metallicity and solar-metallicity bin, the super solar metallicity bin shows a clear non-linear trend, especially at the youngest ages. This figure suggests that the Na abundances of the high-Ia and low-Ia stars prefer a different model and cannot both be explained by a single linear model.

\textbf{3. Spatially uniform IMF:} The fiducial model assumes that the integrated supernovae yields do not depend on spatial or orbital parameters. If different CCSN or SNIa populations enrich with varying strengths across the Galactic disk, such as through a radially dependent IMF, the fiducial model could not fully capture the radial abundance trends from these nucleosynthetic sources. Some evidence has been found for a more top-heavy IMF in the inner Galaxy than solar neighborhood \citep{ballero2007, grieco2012, grieco2015, horta2022}. A simplistic comparison of the inner and outer Galaxy in \citet{griffith2021a} does not show clear correlations with predicted residuals from a varying IMF slope.

\textbf{4. MW Disk Stars:} Another underlying assumption of the fiducial model is that the stars included in our sample are MW disk stars. In practice, accreted structures lie within the MW halo \citep{horta2023}, and some of these stars have likely contaminated our sample. The largest halo substructure, GSE \citep{belokurov2018, helmi2018, mackereth2019}, is known to have anomalous C+N, Na, and Al abundances, relative to the MW disk \citep{hasselquist2024}. Accreted substructures could be responsible for non-zero residual abundances in the high-$\Zm$ stars, though our metallicity cut should exclude most of the GSE population.

\textbf{5. Non-uniform data density:} As discussed in Section~\ref{sec:data}, we employ APOGEE DR17 abundances with quality and stellar population cuts. This results in a sample of $\N$ stars distributed non-uniformly across the Galactic disk.(Figure~\ref{fig:density_maps.pdf}). KPM optimizes the fit parameters to the underlying stellar population. If the best-fit KPM model varies with spatial location, a fit to the whole disk will best optimize the densest data region, in this case, the mid-plane solar neighborhood. This underlying model assumption does not cause the observed residual trends but does make their absolute values and zero-points difficult to interpret. 

Each limitation of KPM described here may contribute to the residual structure that correlates with $\Rg$, $\Zm$, and age. It is unlikely that a single assumption alone is fully causing the observed variance. We hope to improve the flexibility of KPM in future work, exploring the detailed impact of the fiducial model assumptions.

\section{Discussion}\label{sec:discussion}

In this article, we present abundance residuals away from a fiducial two-parameter nucleosynthetic model (KPM) and their coherent and intricate structure with respect to guiding radius, vertical action, and age dimensions.
Our sample of $\N$ stars (described in Section~\ref{sec:data}) are observed by APOGEE DR17 and span the Galactic disk, from the bulge to the outskirts ($R \sim 16$~kpc). In the fiducial model (Section~\ref{sec:fitting_kpm}), KPM describes the $\logg$ calibrated abundances (S24) of 15 elements as the sum of two processes---one Mg-like process tracking CCSN enrichment and one Fe-like process tracking SNIa enrichment. We calculate residual [X/H] abundances away from the KPM predictions, finding that while most elements in most stars are well predicted with $\overline{|\Delta [\rm X/H]|}<0.05$~dex for all but Na (0.06) and Ce (0.08), there are meaningful deviations away from the reported abundances.

Similar two-parameter models have been thoroughly explored in recent literature \citep[e.g.,][G24, S24]{weinberg2019, griffith2021a}. Notably, these works find that while the dilution \citep{chiappini1997, horta2021}, IMF \citep{ballero2007, grieco2012, grieco2015, horta2022}, SFE \citep{rybizki2017, andrews2017}, and other conditions of the star-forming environment may change radially and temporally, a single two-parameter model can describe the APOGEE disk star abundances at all radii and ages to $\sim 0.05$ dex.
This universality suggests that the MW abundances are (at least approximately) low dimensional and that, at large amplitude, two-parameter models probe the underlying integrated nucleosynthetic yields that enrich the disk, not the local star formation history.

Here, we characterize the intricate behavior of the abundance residuals at smaller amplitudes to test if additional information and dimensionality exist, or if residuals are a product of abundance scatter and systematics. While cross-element residual correlations can be useful in such analysis, they are also very susceptible to abundance systematics with stellar parameters like $\teff$ and $\logg$ that impact multiple elements in similar ways \citep{ting2022}. Searching for residual correlations with non-nucleosynthetic parameters, such as orbits and actions, helps us reduce the impact of remaining systematics. Adjacent works, done by \citet{hawkins2023, hackshaw2024}, have found that when the average radial metallicity trend is removed, low amplitude azimuthal variations resembling spiral arms are detected.

To identify KPM residual correlations with orbits, actions, and ages, we create maps of the average residual abundances in $\Zm$ vs. $\Rg$ (Figure~\ref{fig:resid_map_Z_R}) and age vs. $\Rg$ (Figures~\ref{fig:resid_map_age_R}) and describe their structures in Section~\ref{sec:resids}. These maps show important and real trends in the distribution of detailed abundance ratios, even though their quantitative amplitudes are very low.
Indeed, the trends we find are at or below the few $10^{-2}$~dex level, even though the input data makes measurements with uncertainties comparable to or larger than the effects we present.
This shows the value and power of having large samples of stars with many measured abundances, and suggests that surveys that observe a greater variety of elements (e.g. GALAH, \citealp{buder2024}) or much larger numbers of stars (e.g. LAMOST, \citealp{lamost}) might find even more trends and effects.
It also suggests that it would be valuable to make even more precise measurements of stellar element abundances, if future developments with new data make that possible.

We observe complex, small amplitude residual trends in all three dimensions away from the fiducial KPM model. The residual abundance structure represent departures from the average KPM prediction. While their sign and zero-point are interesting, these characteristics depend on the dominant population in the underlying model, mid-plane solar neighborhood stars in this work.
Thus we focus on the residuals' qualitative structure. In Figures~\ref{fig:resid_map_Z_R} and~\ref{fig:resid_map_age_R}, we observe residual structure and gradients that resemble components of the Galactic disk, especially in the high-Ia stellar population. This structure has low dimensionality, with three PCA eigenvectors describing 90\% of the residual variance (Section~\ref{subsec:PCA}). 

In $\Zm$ vs. $\Rg$ space, residual Si, Na, Mn, and Co abundances (and the second PCA eigenvector) in the high-Ia population resemble the chemical thin and thick disks, with residual gradients in $\Zm$ that are strongest at $\Zm \lesssim 0.5$ kpc and $\Rg \lesssim 10$ kpc. The kinematic thin and thick disks have been well characterized in the recent decade with Gaia data\citep[e.g.,][]{gaia2023, chandra2024}. Though our chemically selected high-Ia population is dominated by thin disk stars, chemical and kinematic definitions of the thin disk differ, so we expect some kinematic thick disk stars to contaminate our high-Ia sample \citep[e.g.,][]{hayden2017}. The exact origin of the thin and thick disks is debated, but they likely have unique formation histories, with the high-Ia sequence formed by a dilution event \citep[e.g.,][]{chiappini1997, spitoni2019, spitoni2024} and radial migration \citep[e.g.,][]{schonrich2009, minchev2013}. While there is less structure in the low-Ia disk, the Si, S, K, and Ce residuals show a trend with $\Rg$, such that the inner galaxy prefers a slightly different model than the solar neighborhood. This is echoed in the high-Ia disk for S, K, and Ce (and the third and fifth PCA eigenvectors). These residual trends suggest a radial abundance gradient in the high-Ia disk, albeit small in amplitude. In age vs. $\Rg$, we find that the residual abundances tend to be positive in the young stars, especially at $\Rg \lesssim 8$ kpc. Conversely, the residuals of older stars (age $\gtrsim 6$ Gyrs) are preferentially negative. These trends project into $\Rg$ and $\Zm$ as a flared outer disk \citep{hayden2015, bovy2016}.

The fact that the residual abundance structure resembles the kinematic components of the MW disk is remarkable. While we don't have a causal explanation of this structure, its existence shows that we are sensitive to the conditions of the star-forming environment at the time and location of stellar birth. We find that some residual structures, particularly in C+N, Ni, and Mn, cannot be explained by an additional nucleosynthetic source but can be partially explained by a simple dilution model  (Section~\ref{sec:causes_resids}). This is expected if the two-process model traces an underlying chemical equilibrium state that was perturbed by the dilution \citep[e.g.,][]{johnson2024}. Our findings suggest that a star's abundances depend not only on its current [Mg/H] and [Fe/H], but also on how it arrived there. This information is encoded in the other element abundances at small amplitudes, proving that more than two elements matter in describing the nucleosynthesis of the Galactic disk.

We predict that the addition of model flexibility to capture complex metallicity dependence, outflows, IMF variations, and other star formation parameters could further explain the structured small amplitude residuals found here. We intend to explore the impact of these parameters in future versions of KPM. We note that in this work, we are analyzing abundance features at or below the level of APOGEE's abundance precision. Large population analysis, where uncertain values from individual stars can be averaged to reduce the error and increase the statistical significance, is key to understanding the intricate details of Galactic chemical evolution from an observational perspective.

\section*{Acknowledgments}   

We thank Rachel Beaton, Chris Hayes, Alexander Stone-Martinez, Melissa Ness, Hans-Walter Rix, and the Darling research group at CU Boulder for valuable discussions and help on this project. 

E.J.G. is supported by an NSF Astronomy and Astrophysics Postdoctoral Fellowship under award AST-2202135. The Flatiron Institute is a division of the Simons Foundation.

Funding for the Sloan Digital Sky Survey IV has been provided by the Alfred P. Sloan Foundation, the U.S. Department of Energy Office of Science, and the Participating Institutions. 

SDSS-IV acknowledges support and resources from the Center for High Performance Computing at the University of Utah. The SDSS website is www.sdss4.org.

SDSS-IV is managed by the Astrophysical Research Consortium for the Participating Institutions of the SDSS Collaboration including the Brazilian Participation Group, the Carnegie Institution for Science, Carnegie Mellon University, Center for Astrophysics | Harvard \& Smithsonian, the Chilean Participation Group, the French Participation Group, Instituto de Astrof\'isica de Canarias, The Johns Hopkins University, Kavli Institute for the Physics and Mathematics of the Universe (IPMU) / University of Tokyo, the Korean Participation Group, Lawrence Berkeley National Laboratory, Leibniz Institut f\"ur Astrophysik Potsdam (AIP),  Max-Planck-Institut f\"ur Astronomie (MPIA Heidelberg), Max-Planck-Institut f\"ur Astrophysik (MPA Garching), Max-Planck-Institut f\"ur Extraterrestrische Physik (MPE), National Astronomical Observatories of China, New Mexico State University, New York University, University of Notre Dame, Observat\'ario Nacional / MCTI, The Ohio State University, Pennsylvania State University, Shanghai Astronomical Observatory, United Kingdom Participation Group, Universidad Nacional Aut\'onoma de M\'exico, University of Arizona, University of Colorado Boulder, University of Oxford, University of Portsmouth, University of Utah, University of Virginia, University of Washington, University of Wisconsin, Vanderbilt University, and Yale University.

\software{
Astropy \citep{astropy2013, astropy2018, astropy2022},
cmastro (\url{cmastro.readthedocs.io}),
gala \citep{Price-Whelan:2017}, 
jax \citep{jax},
Matplotlib \citep{hunter2007}, 
NumPy \citep{harris2020}, 
Pandas \citep{pandasa, pandasb}, 
scikit-learn \citep{scikit-learn}
}

\bibliography{sample631}{}

\begin{thebibliography}{}
\expandafter\ifx\csname natexlab\endcsname\relax\def\natexlab#1{#1}\fi
\providecommand{\url}[1]{\href{#1}{#1}}
\providecommand{\dodoi}[1]{doi:~\href{http://doi.org/#1}{\nolinkurl{#1}}}
\providecommand{\doeprint}[1]{\href{http://ascl.net/#1}{\nolinkurl{http://ascl.net/#1}}}
\providecommand{\doarXiv}[1]{\href{https://arxiv.org/abs/#1}{\nolinkurl{https://arxiv.org/abs/#1}}}

\bibitem[{{Abdurro'uf} {et~al.}(2022){Abdurro'uf}, {Accetta}, {Aerts}, {Silva Aguirre}, {Ahumada}, {Ajgaonkar}, {Filiz Ak}, {Alam}, {Allende Prieto}, {Almeida}, {Anders}, {Anderson}, {Andrews}, {Anguiano}, {Aquino-Ort{\'\i}z}, {Arag{\'o}n-Salamanca}, {Argudo-Fern{\'a}ndez}, {Ata}, {Aubert}, {Avila-Reese}, {Badenes}, {Barb{\'a}}, {Barger}, {Barrera-Ballesteros}, {Beaton}, {Beers}, {Belfiore}, {Bender}, {Bernardi}, {Bershady}, {Beutler}, {Bidin}, {Bird}, {Bizyaev}, {Blanc}, {Blanton}, {Boardman}, {Bolton}, {Boquien}, {Borissova}, {Bovy}, {Brandt}, {Brown}, {Brownstein}, {Brusa}, {Buchner}, {Bundy}, {Burchett}, {Bureau}, {Burgasser}, {Cabang}, {Campbell}, {Cappellari}, {Carlberg}, {Wanderley}, {Carrera}, {Cash}, {Chen}, {Chen}, {Cherinka}, {Chiappini}, {Choi}, {Chojnowski}, {Chung}, {Clerc}, {Cohen}, {Comerford}, {Comparat}, {da Costa}, {Covey}, {Crane}, {Cruz-Gonzalez}, {Culhane}, {Cunha}, {Dai}, {Damke}, {Darling}, {Davidson}, {Davies}, {Dawson}, {De Lee}, {Diamond-Stanic}, {Cano-D{\'\i}az}, {S{\'a}nchez},
  {Donor}, {Duckworth}, {Dwelly}, {Eisenstein}, {Elsworth}, {Emsellem}, {Eracleous}, {Escoffier}, {Fan}, {Farr}, {Feng}, {Fern{\'a}ndez-Trincado}, {Feuillet}, {Filipp}, {Fillingham}, {Frinchaboy}, {Fromenteau}, {Galbany}, {Garc{\'\i}a}, {Garc{\'\i}a-Hern{\'a}ndez}, {Ge}, {Geisler}, {Gelfand}, {G{\'e}ron}, {Gibson}, {Goddy}, {Godoy-Rivera}, {Grabowski}, {Green}, {Greener}, {Grier}, {Griffith}, {Guo}, {Guy}, {Hadjara}, {Harding}, {Hasselquist}, {Hayes}, {Hearty}, {Hern{\'a}ndez}, {Hill}, {Hogg}, {Holtzman}, {Horta}, {Hsieh}, {Hsu}, {Hsu}, {Huber}, {Huertas-Company}, {Hutchinson}, {Hwang}, {Ibarra-Medel}, {Chitham}, {Ilha}, {Imig}, {Jaekle}, {Jayasinghe}, {Ji}, {Johnson}, {Jones}, {J{\"o}nsson}, {Katkov}, {Khalatyan}, {Kinemuchi}, {Kisku}, {Knapen}, {Kneib}, {Kollmeier}, {Kong}, {Kounkel}, {Kreckel}, {Krishnarao}, {Lacerna}, {Lane}, {Langgin}, {Lavender}, {Law}, {Lazarz}, {Leung}, {Leung}, {Lewis}, {Li}, {Li}, {Lian}, {Liang}, {Lin}, {Lin}, {Lin}, {Lintott}, {Long}, {Longa-Pe{\~n}a}, {L{\'o}pez-Cob{\'a}}, {Lu},
  {Lundgren}, {Luo}, {Mackereth}, {de la Macorra}, {Mahadevan}, {Majewski}, {Manchado}, {Mandeville}, {Maraston}, {Margalef-Bentabol}, {Masseron}, {Masters}, {Mathur}, {McDermid}, {Mckay}, {Merloni}, {Merrifield}, {Meszaros}, {Miglio}, {Di Mille}, {Minniti}, {Minsley}, {Monachesi}, {Moon}, {Mosser}, {Mulchaey}, {Muna}, {Mu{\~n}oz}, {Myers}, {Myers}, {Nadathur}, {Nair}, {Nandra}, {Neumann}, {Newman}, {Nidever}, {Nikakhtar}, {Nitschelm}, {O'Connell}, {Garma-Oehmichen}, {Luan Souza de Oliveira}, {Olney}, {Oravetz}, {Ortigoza-Urdaneta}, {Osorio}, {Otter}, {Pace}, {Padilla}, {Pan}, {Pan}, {Parikh}, {Parker}, {Peirani}, {Pe{\~n}a Ram{\'\i}rez}, {Penny}, {Percival}, {Perez-Fournon}, {Pinsonneault}, {Poidevin}, {Poovelil}, {Price-Whelan}, {B{\'a}rbara de Andrade Queiroz}, {Raddick}, {Ray}, {Rembold}, {Riddle}, {Riffel}, {Riffel}, {Rix}, {Robin}, {Rodr{\'\i}guez-Puebla}, {Roman-Lopes}, {Rom{\'a}n-Z{\'u}{\~n}iga}, {Rose}, {Ross}, {Rossi}, {Rubin}, {Salvato}, {S{\'a}nchez}, {S{\'a}nchez-Gallego}, {Sanderson}, {Santana
  Rojas}, {Sarceno}, {Sarmiento}, {Sayres}, {Sazonova}, {Schaefer}, {Schiavon}, {Schlegel}, {Schneider}, {Schultheis}, {Schwope}, {Serenelli}, {Serna}, {Shao}, {Shapiro}, {Sharma}, {Shen}, {Shetrone}, {Shu}, {Simon}, {Skrutskie}, {Smethurst}, {Smith}, {Sobeck}, {Spoo}, {Sprague}, {Stark}, {Stassun}, {Steinmetz}, {Stello}, {Stone-Martinez}, {Storchi-Bergmann}, {Stringfellow}, {Stutz}, {Su}, {Taghizadeh-Popp}, {Talbot}, {Tayar}, {Telles}, {Teske}, {Thakar}, {Theissen}, {Tkachenko}, {Thomas}, {Tojeiro}, {Hernandez Toledo}, {Troup}, {Trump}, {Trussler}, {Turner}, {Tuttle}, {Unda-Sanzana}, {V{\'a}zquez-Mata}, {Valentini}, {Valenzuela}, {Vargas-Gonz{\'a}lez}, {Vargas-Maga{\~n}a}, {Alfaro}, {Villanova}, {Vincenzo}, {Wake}, {Warfield}, {Washington}, {Weaver}, {Weijmans}, {Weinberg}, {Weiss}, {Westfall}, {Wild}, {Wilde}, {Wilson}, {Wilson}, {Wilson}, {Wolf}, {Wood-Vasey}, {Yan}, {Zamora}, {Zasowski}, {Zhang}, {Zhao}, {Zheng}, {Zheng}, \& {Zhu}}]{abdurrouf2022}
{Abdurro'uf}, {Accetta}, K., {Aerts}, C., {et~al.} 2022, \apjs, 259, 35, \dodoi{10.3847/1538-4365/ac4414}

\bibitem[{{Andrews} {et~al.}(2017){Andrews}, {Weinberg}, {Sch{\"o}nrich}, \& {Johnson}}]{andrews2017}
{Andrews}, B.~H., {Weinberg}, D.~H., {Sch{\"o}nrich}, R., \& {Johnson}, J.~A. 2017, \apj, 835, 224, \dodoi{10.3847/1538-4357/835/2/224}

\bibitem[{{Arlandini} {et~al.}(1999){Arlandini}, {K{\"a}ppeler}, {Wisshak}, {Gallino}, {Lugaro}, {Busso}, \& {Straniero}}]{arlandini1999}
{Arlandini}, C., {K{\"a}ppeler}, F., {Wisshak}, K., {et~al.} 1999, \apj, 525, 886, \dodoi{10.1086/307938}

\bibitem[{{Astropy Collaboration} {et~al.}(2013){Astropy Collaboration}, {Robitaille}, {Tollerud}, {Greenfield}, {Droettboom}, {Bray}, {Aldcroft}, {Davis}, {Ginsburg}, {Price-Whelan}, {Kerzendorf}, {Conley}, {Crighton}, {Barbary}, {Muna}, {Ferguson}, {Grollier}, {Parikh}, {Nair}, {Unther}, {Deil}, {Woillez}, {Conseil}, {Kramer}, {Turner}, {Singer}, {Fox}, {Weaver}, {Zabalza}, {Edwards}, {Azalee Bostroem}, {Burke}, {Casey}, {Crawford}, {Dencheva}, {Ely}, {Jenness}, {Labrie}, {Lim}, {Pierfederici}, {Pontzen}, {Ptak}, {Refsdal}, {Servillat}, \& {Streicher}}]{astropy2013}
{Astropy Collaboration}, {Robitaille}, T.~P., {Tollerud}, E.~J., {et~al.} 2013, \aap, 558, A33, \dodoi{10.1051/0004-6361/201322068}

\bibitem[{{Astropy Collaboration} {et~al.}(2018){Astropy Collaboration}, {Price-Whelan}, {Sip{\H{o}}cz}, {G{\"u}nther}, {Lim}, {Crawford}, {Conseil}, {Shupe}, {Craig}, {Dencheva}, {Ginsburg}, {Vand erPlas}, {Bradley}, {P{\'e}rez-Su{\'a}rez}, {de Val-Borro}, {Aldcroft}, {Cruz}, {Robitaille}, {Tollerud}, {Ardelean}, {Babej}, {Bach}, {Bachetti}, {Bakanov}, {Bamford}, {Barentsen}, {Barmby}, {Baumbach}, {Berry}, {Biscani}, {Boquien}, {Bostroem}, {Bouma}, {Brammer}, {Bray}, {Breytenbach}, {Buddelmeijer}, {Burke}, {Calderone}, {Cano Rodr{\'\i}guez}, {Cara}, {Cardoso}, {Cheedella}, {Copin}, {Corrales}, {Crichton}, {D'Avella}, {Deil}, {Depagne}, {Dietrich}, {Donath}, {Droettboom}, {Earl}, {Erben}, {Fabbro}, {Ferreira}, {Finethy}, {Fox}, {Garrison}, {Gibbons}, {Goldstein}, {Gommers}, {Greco}, {Greenfield}, {Groener}, {Grollier}, {Hagen}, {Hirst}, {Homeier}, {Horton}, {Hosseinzadeh}, {Hu}, {Hunkeler}, {Ivezi{\'c}}, {Jain}, {Jenness}, {Kanarek}, {Kendrew}, {Kern}, {Kerzendorf}, {Khvalko}, {King}, {Kirkby}, {Kulkarni},
  {Kumar}, {Lee}, {Lenz}, {Littlefair}, {Ma}, {Macleod}, {Mastropietro}, {McCully}, {Montagnac}, {Morris}, {Mueller}, {Mumford}, {Muna}, {Murphy}, {Nelson}, {Nguyen}, {Ninan}, {N{\"o}the}, {Ogaz}, {Oh}, {Parejko}, {Parley}, {Pascual}, {Patil}, {Patil}, {Plunkett}, {Prochaska}, {Rastogi}, {Reddy Janga}, {Sabater}, {Sakurikar}, {Seifert}, {Sherbert}, {Sherwood-Taylor}, {Shih}, {Sick}, {Silbiger}, {Singanamalla}, {Singer}, {Sladen}, {Sooley}, {Sornarajah}, {Streicher}, {Teuben}, {Thomas}, {Tremblay}, {Turner}, {Terr{\'o}n}, {van Kerkwijk}, {de la Vega}, {Watkins}, {Weaver}, {Whitmore}, {Woillez}, {Zabalza}, \& {Astropy Contributors}}]{astropy2018}
{Astropy Collaboration}, {Price-Whelan}, A.~M., {Sip{\H{o}}cz}, B.~M., {et~al.} 2018, \aj, 156, 123, \dodoi{10.3847/1538-3881/aabc4f}

\bibitem[{{Astropy Collaboration} {et~al.}(2022){Astropy Collaboration}, {Price-Whelan}, {Lim}, {Earl}, {Starkman}, {Bradley}, {Shupe}, {Patil}, {Corrales}, {Brasseur}, {N{"o}the}, {Donath}, {Tollerud}, {Morris}, {Ginsburg}, {Vaher}, {Weaver}, {Tocknell}, {Jamieson}, {van Kerkwijk}, {Robitaille}, {Merry}, {Bachetti}, {G{"u}nther}, {Aldcroft}, {Alvarado-Montes}, {Archibald}, {B{'o}di}, {Bapat}, {Barentsen}, {Baz{'a}n}, {Biswas}, {Boquien}, {Burke}, {Cara}, {Cara}, {Conroy}, {Conseil}, {Craig}, {Cross}, {Cruz}, {D'Eugenio}, {Dencheva}, {Devillepoix}, {Dietrich}, {Eigenbrot}, {Erben}, {Ferreira}, {Foreman-Mackey}, {Fox}, {Freij}, {Garg}, {Geda}, {Glattly}, {Gondhalekar}, {Gordon}, {Grant}, {Greenfield}, {Groener}, {Guest}, {Gurovich}, {Handberg}, {Hart}, {Hatfield-Dodds}, {Homeier}, {Hosseinzadeh}, {Jenness}, {Jones}, {Joseph}, {Kalmbach}, {Karamehmetoglu}, {Ka{l}uszy{'n}ski}, {Kelley}, {Kern}, {Kerzendorf}, {Koch}, {Kulumani}, {Lee}, {Ly}, {Ma}, {MacBride}, {Maljaars}, {Muna}, {Murphy}, {Norman}, {O'Steen},
  {Oman}, {Pacifici}, {Pascual}, {Pascual-Granado}, {Patil}, {Perren}, {Pickering}, {Rastogi}, {Roulston}, {Ryan}, {Rykoff}, {Sabater}, {Sakurikar}, {Salgado}, {Sanghi}, {Saunders}, {Savchenko}, {Schwardt}, {Seifert-Eckert}, {Shih}, {Jain}, {Shukla}, {Sick}, {Simpson}, {Singanamalla}, {Singer}, {Singhal}, {Sinha}, {Sip{H{o}}cz}, {Spitler}, {Stansby}, {Streicher}, {{{S}}umak}, {Swinbank}, {Taranu}, {Tewary}, {Tremblay}, {Val-Borro}, {Van Kooten}, {Vasovi{'c}}, {Verma}, {de Miranda Cardoso}, {Williams}, {Wilson}, {Winkel}, {Wood-Vasey}, {Xue}, {Yoachim}, {Zhang}, {Zonca}, \& {Astropy Project Contributors}}]{astropy2022}
{Astropy Collaboration}, {Price-Whelan}, A.~M., {Lim}, P.~L., {et~al.} 2022, ApJ, 935, 167, \dodoi{10.3847/1538-4357/ac7c74}

\bibitem[{{Ballero} {et~al.}(2007){Ballero}, {Kroupa}, \& {Matteucci}}]{ballero2007}
{Ballero}, S.~K., {Kroupa}, P., \& {Matteucci}, F. 2007, \aap, 467, 117, \dodoi{10.1051/0004-6361:20066786}

\bibitem[{{Beaton} {et~al.}(2021){Beaton}, {Oelkers}, {Hayes}, {Covey}, {Chojnowski}, {De Lee}, {Sobeck}, {Majewski}, {Cohen}, {Fern{\'a}ndez-Trincado}, {Longa-Pe{\~n}a}, {O'Connell}, {Santana}, {Stringfellow}, {Zasowski}, {Aerts}, {Anguiano}, {Bender}, {Ca{\~n}as}, {Cunha}, {Donor}, {Fleming}, {Frinchaboy}, {Feuillet}, {Harding}, {Hasselquist}, {Holtzman}, {Johnson}, {Kollmeier}, {Kounkel}, {Mahadevan}, {Price-Whelan}, {Rojas-Arriagada}, {Rom{\'a}n-Z{\'u}{\~n}iga}, {Schlafly}, {Schultheis}, {Shetrone}, {Simon}, {Stassun}, {Stutz}, {Tayar}, {Teske}, {Tkachenko}, {Troup}, {Albareti}, {Bizyaev}, {Bovy}, {Burgasser}, {Comparat}, {Downes}, {Geisler}, {Inno}, {Manchado}, {Ness}, {Pinsonneault}, {Prada}, {Roman-Lopes}, {Simonian}, {Smith}, {Yan}, \& {Zamora}}]{beaton2021}
{Beaton}, R.~L., {Oelkers}, R.~J., {Hayes}, C.~R., {et~al.} 2021, \aj, 162, 302, \dodoi{10.3847/1538-3881/ac260c}

\bibitem[{{Belokurov} {et~al.}(2018){Belokurov}, {Erkal}, {Evans}, {Koposov}, \& {Deason}}]{belokurov2018}
{Belokurov}, V., {Erkal}, D., {Evans}, N.~W., {Koposov}, S.~E., \& {Deason}, A.~J. 2018, \mnras, 478, 611, \dodoi{10.1093/mnras/sty982}

\bibitem[{{Bennett} \& {Bovy}(2019)}]{Bennett:2019}
{Bennett}, M., \& {Bovy}, J. 2019, \mnras, 482, 1417, \dodoi{10.1093/mnras/sty2813}

\bibitem[{{Bensby} {et~al.}(2013){Bensby}, {Yee}, {Feltzing}, {Johnson}, {Gould}, {Cohen}, {Asplund}, {Mel{\'e}ndez}, {Lucatello}, {Han}, {Thompson}, {Gal-Yam}, {Udalski}, {Bennett}, {Bond}, {Kohei}, {Sumi}, {Suzuki}, {Suzuki}, {Takino}, {Tristram}, {Yamai}, \& {Yonehara}}]{bensby2013}
{Bensby}, T., {Yee}, J.~C., {Feltzing}, S., {et~al.} 2013, \aap, 549, A147, \dodoi{10.1051/0004-6361/201220678}

\bibitem[{{Blanton} {et~al.}(2017){Blanton}, {Bershady}, {Abolfathi}, {Albareti}, {Allende Prieto}, {Almeida}, {Alonso-Garc{\'\i}a}, {Anders}, {Anderson}, {Andrews}, {Aquino-Ort{\'\i}z}, {Arag{\'o}n-Salamanca}, {Argudo-Fern{\'a}ndez}, {Armengaud}, {Aubourg}, {Avila-Reese}, {Badenes}, {Bailey}, {Barger}, {Barrera-Ballesteros}, {Bartosz}, {Bates}, {Baumgarten}, {Bautista}, {Beaton}, {Beers}, {Belfiore}, {Bender}, {Berlind}, {Bernardi}, {Beutler}, {Bird}, {Bizyaev}, {Blanc}, {Blomqvist}, {Bolton}, {Boquien}, {Borissova}, {van den Bosch}, {Bovy}, {Brandt}, {Brinkmann}, {Brownstein}, {Bundy}, {Burgasser}, {Burtin}, {Busca}, {Cappellari}, {Delgado Carigi}, {Carlberg}, {Carnero Rosell}, {Carrera}, {Chanover}, {Cherinka}, {Cheung}, {G{\'o}mez Maqueo Chew}, {Chiappini}, {Choi}, {Chojnowski}, {Chuang}, {Chung}, {Cirolini}, {Clerc}, {Cohen}, {Comparat}, {da Costa}, {Cousinou}, {Covey}, {Crane}, {Croft}, {Cruz-Gonzalez}, {Garrido Cuadra}, {Cunha}, {Damke}, {Darling}, {Davies}, {Dawson}, {de la Macorra}, {Dell'Agli}, {De
  Lee}, {Delubac}, {Di Mille}, {Diamond-Stanic}, {Cano-D{\'\i}az}, {Donor}, {Downes}, {Drory}, {du Mas des Bourboux}, {Duckworth}, {Dwelly}, {Dyer}, {Ebelke}, {Eigenbrot}, {Eisenstein}, {Emsellem}, {Eracleous}, {Escoffier}, {Evans}, {Fan}, {Fern{\'a}ndez-Alvar}, {Fernandez-Trincado}, {Feuillet}, {Finoguenov}, {Fleming}, {Font-Ribera}, {Fredrickson}, {Freischlad}, {Frinchaboy}, {Fuentes}, {Galbany}, {Garcia-Dias}, {Garc{\'\i}a-Hern{\'a}ndez}, {Gaulme}, {Geisler}, {Gelfand}, {Gil-Mar{\'\i}n}, {Gillespie}, {Goddard}, {Gonzalez-Perez}, {Grabowski}, {Green}, {Grier}, {Gunn}, {Guo}, {Guy}, {Hagen}, {Hahn}, {Hall}, {Harding}, {Hasselquist}, {Hawley}, {Hearty}, {Gonzalez Hern{\'a}ndez}, {Ho}, {Hogg}, {Holley-Bockelmann}, {Holtzman}, {Holzer}, {Huehnerhoff}, {Hutchinson}, {Hwang}, {Ibarra-Medel}, {da Silva Ilha}, {Ivans}, {Ivory}, {Jackson}, {Jensen}, {Johnson}, {Jones}, {J{\"o}nsson}, {Jullo}, {Kamble}, {Kinemuchi}, {Kirkby}, {Kitaura}, {Klaene}, {Knapp}, {Kneib}, {Kollmeier}, {Lacerna}, {Lane}, {Lang}, {Law},
  {Lazarz}, {Lee}, {Le Goff}, {Liang}, {Li}, {Li}, {Lian}, {Lima}, {Lin}, {Lin}, {Bertran de Lis}, {Liu}, {de Icaza Lizaola}, {Long}, {Lucatello}, {Lundgren}, {MacDonald}, {Deconto Machado}, {MacLeod}, {Mahadevan}, {Geimba Maia}, {Maiolino}, {Majewski}, {Malanushenko}, {Malanushenko}, {Manchado}, {Mao}, {Maraston}, {Marques-Chaves}, {Masseron}, {Masters}, {McBride}, {McDermid}, {McGrath}, {McGreer}, {Medina Pe{\~n}a}, {Melendez}, {Merloni}, {Merrifield}, {Meszaros}, {Meza}, {Minchev}, {Minniti}, {Miyaji}, {More}, {Mulchaey}, {M{\"u}ller-S{\'a}nchez}, {Muna}, {Munoz}, {Myers}, {Nair}, {Nandra}, {Correa do Nascimento}, {Negrete}, {Ness}, {Newman}, {Nichol}, {Nidever}, {Nitschelm}, {Ntelis}, {O'Connell}, {Oelkers}, {Oravetz}, {Oravetz}, {Pace}, {Padilla}, {Palanque-Delabrouille}, {Alonso Palicio}, {Pan}, {Parejko}, {Parikh}, {P{\^a}ris}, {Park}, {Patten}, {Peirani}, {Pellejero-Ibanez}, {Penny}, {Percival}, {Perez-Fournon}, {Petitjean}, {Pieri}, {Pinsonneault}, {Pisani}, {Poleski}, {Prada}, {Prakash}, {Queiroz},
  {Raddick}, {Raichoor}, {Barboza Rembold}, {Richstein}, {Riffel}, {Riffel}, {Rix}, {Robin}, {Rockosi}, {Rodr{\'\i}guez-Torres}, {Roman-Lopes}, {Rom{\'a}n-Z{\'u}{\~n}iga}, {Rosado}, {Ross}, {Rossi}, {Ruan}, {Ruggeri}, {Rykoff}, {Salazar-Albornoz}, {Salvato}, {S{\'a}nchez}, {Aguado}, {S{\'a}nchez-Gallego}, {Santana}, {Santiago}, {Sayres}, {Schiavon}, {da Silva Schimoia}, {Schlafly}, {Schlegel}, {Schneider}, {Schultheis}, {Schuster}, {Schwope}, {Seo}, {Shao}, {Shen}, {Shetrone}, {Shull}, {Simon}, {Skinner}, {Skrutskie}, {Slosar}, {Smith}, {Sobeck}, {Sobreira}, {Somers}, {Souto}, {Stark}, {Stassun}, {Stauffer}, {Steinmetz}, {Storchi-Bergmann}, {Streblyanska}, {Stringfellow}, {Su{\'a}rez}, {Sun}, {Suzuki}, {Szigeti}, {Taghizadeh-Popp}, {Tang}, {Tao}, {Tayar}, {Tembe}, {Teske}, {Thakar}, {Thomas}, {Thompson}, {Tinker}, {Tissera}, {Tojeiro}, {Hernandez Toledo}, {de la Torre}, {Tremonti}, {Troup}, {Valenzuela}, {Martinez Valpuesta}, {Vargas-Gonz{\'a}lez}, {Vargas-Maga{\~n}a}, {Vazquez}, {Villanova}, {Vivek}, {Vogt},
  {Wake}, {Walterbos}, {Wang}, {Weaver}, {Weijmans}, {Weinberg}, {Westfall}, {Whelan}, {Wild}, {Wilson}, {Wood-Vasey}, {Wylezalek}, {Xiao}, {Yan}, {Yang}, {Ybarra}, {Y{\`e}che}, {Zakamska}, {Zamora}, {Zarrouk}, {Zasowski}, {Zhang}, {Zhao}, {Zheng}, {Zheng}, {Zhou}, {Zhou}, {Zhu}, {Zoccali}, \& {Zou}}]{blanton2017}
{Blanton}, M.~R., {Bershady}, M.~A., {Abolfathi}, B., {et~al.} 2017, \aj, 154, 28, \dodoi{10.3847/1538-3881/aa7567}

\bibitem[{{Bovy} \& {Rix}(2013)}]{bovy2016}
{Bovy}, J., \& {Rix}, H.-W. 2013, \apj, 779, 115, \dodoi{10.1088/0004-637X/779/2/115}

\bibitem[{{Bowen} \& {Vaughan}(1973)}]{bowen1973}
{Bowen}, I.~S., \& {Vaughan}, A.~H., J. 1973, \ao, 12, 1430, \dodoi{10.1364/AO.12.001430}

\bibitem[{Bradbury {et~al.}(2018)Bradbury, Frostig, Hawkins, Johnson, Leary, Maclaurin, Necula, Paszke, Vander{P}las, Wanderman-{M}ilne, \& Zhang}]{jax}
Bradbury, J., Frostig, R., Hawkins, P., {et~al.} 2018, {JAX}: composable transformations of {P}ython+{N}um{P}y programs, 0.3.13.
\newblock \url{http://github.com/jax-ml/jax}

\bibitem[{{Buder} {et~al.}(2024){Buder}, {Kos}, {Wang}, {McKenzie}, {Howell}, {Martell}, {Hayden}, {Zucker}, {Nordlander}, {Montet}, {Traven}, {Bland-Hawthorn}, {De Silva}, {Freeman}, {Lewis}, {Lind}, {Sharma}, {Simpson}, {Stello}, {Zwitter}, {Amarsi}, {Armstrong}, {Banks}, {Beavis}, {Beeson}, {Chen}, {Ciuc{\u{a}}}, {Da Costa}, {de Grijs}, {Martin}, {Nataf}, {Ness}, {Rains}, {Scarr}, {Vogrin{\v{c}}i{\v{c}}}, {Wang}, {Wittenmyer}, {Xie}, \& {The GALAH Collaboration}}]{buder2024}
{Buder}, S., {Kos}, J., {Wang}, E.~X., {et~al.} 2024, arXiv e-prints, arXiv:2409.19858, \dodoi{10.48550/arXiv.2409.19858}

\bibitem[{{Casey} {et~al.}(2019){Casey}, {Lattanzio}, {Aleti}, {Dowe}, {Bland-Hawthorn}, {Buder}, {Lewis}, {Martell}, {Nordlander}, {Simpson}, {Sharma}, \& {Zucker}}]{casey2019}
{Casey}, A.~R., {Lattanzio}, J.~C., {Aleti}, A., {et~al.} 2019, \apj, 887, 73, \dodoi{10.3847/1538-4357/ab4fea}

\bibitem[{{Chandra} {et~al.}(2024){Chandra}, {Semenov}, {Rix}, {Conroy}, {Bonaca}, {Naidu}, {Andrae}, {Li}, \& {Hernquist}}]{chandra2024}
{Chandra}, V., {Semenov}, V.~A., {Rix}, H.-W., {et~al.} 2024, \apj, 972, 112, \dodoi{10.3847/1538-4357/ad5b60}

\bibitem[{{Chiappini} {et~al.}(1997){Chiappini}, {Matteucci}, \& {Gratton}}]{chiappini1997}
{Chiappini}, C., {Matteucci}, F., \& {Gratton}, R. 1997, \apj, 477, 765, \dodoi{10.1086/303726}

\bibitem[{{Choi} {et~al.}(2016){Choi}, {Dotter}, {Conroy}, {Cantiello}, {Paxton}, \& {Johnson}}]{choi2016}
{Choi}, J., {Dotter}, A., {Conroy}, C., {et~al.} 2016, \apj, 823, 102, \dodoi{10.3847/0004-637X/823/2/102}

\bibitem[{{Cunha} {et~al.}(2016){Cunha}, {Frinchaboy}, {Souto}, {Thompson}, {Zasowski}, {Allende Prieto}, {Carrera}, {Chiappini}, {Donor}, {Garc{\'\i}a-Hern{\'a}ndez}, {Garc{\'\i}a P{\'e}rez}, {Hayden}, {Holtzman}, {Jackson}, {Johnson}, {Majewski}, {M{\'e}sz{\'a}ros}, {Meyer}, {Nidever}, {O'Connell}, {Schiavon}, {Schultheis}, {Shetrone}, {Simmons}, {Smith}, \& {et al.}}]{cunha2016}
{Cunha}, K., {Frinchaboy}, P.~M., {Souto}, D., {et~al.} 2016, Astronomische Nachrichten, 337, 922, \dodoi{10.1002/asna.201612398}

\bibitem[{{de los Reyes} {et~al.}(2020){de los Reyes}, {Kirby}, {Seitenzahl}, \& {Shen}}]{reyes2020}
{de los Reyes}, M. A.~C., {Kirby}, E.~N., {Seitenzahl}, I.~R., \& {Shen}, K.~J. 2020, \apj, 891, 85, \dodoi{10.3847/1538-4357/ab736f}

\bibitem[{{De Silva} {et~al.}(2015){De Silva}, {Freeman}, {Bland-Hawthorn}, {Martell}, {de Boer}, {Asplund}, {Keller}, {Sharma}, {Zucker}, {Zwitter}, {Anguiano}, {Bacigalupo}, {Bayliss}, {Beavis}, {Bergemann}, {Campbell}, {Cannon}, {Carollo}, {Casagrande}, {Casey}, {Da Costa}, {D'Orazi}, {Dotter}, {Duong}, {Heger}, {Ireland}, {Kafle}, {Kos}, {Lattanzio}, {Lewis}, {Lin}, {Lind}, {Munari}, {Nataf}, {O'Toole}, {Parker}, {Reid}, {Schlesinger}, {Sheinis}, {Simpson}, {Stello}, {Ting}, {Traven}, {Watson}, {Wittenmyer}, {Yong}, \& {{\v{Z}}erjal}}]{desilva2015}
{De Silva}, G.~M., {Freeman}, K.~C., {Bland-Hawthorn}, J., {et~al.} 2015, \mnras, 449, 2604, \dodoi{10.1093/mnras/stv327}

\bibitem[{{Frankel} {et~al.}(2018){Frankel}, {Rix}, {Ting}, {Ness}, \& {Hogg}}]{frankel2018}
{Frankel}, N., {Rix}, H.-W., {Ting}, Y.-S., {Ness}, M., \& {Hogg}, D.~W. 2018, \apj, 865, 96, \dodoi{10.3847/1538-4357/aadba5}

\bibitem[{{Gaia Collaboration} {et~al.}(2023){Gaia Collaboration}, {Vallenari}, {Brown}, {Prusti}, {de Bruijne}, {Arenou}, {Babusiaux}, {Biermann}, {Creevey}, {Ducourant}, {Evans}, {Eyer}, {Guerra}, {Hutton}, {Jordi}, {Klioner}, {Lammers}, {Lindegren}, {Luri}, {Mignard}, {Panem}, {Pourbaix}, {Randich}, {Sartoretti}, {Soubiran}, {Tanga}, {Walton}, {Bailer-Jones}, {Bastian}, {Drimmel}, {Jansen}, {Katz}, {Lattanzi}, {van Leeuwen}, {Bakker}, {Cacciari}, {Casta{\~n}eda}, {De Angeli}, {Fabricius}, {Fouesneau}, {Fr{\'e}mat}, {Galluccio}, {Guerrier}, {Heiter}, {Masana}, {Messineo}, {Mowlavi}, {Nicolas}, {Nienartowicz}, {Pailler}, {Panuzzo}, {Riclet}, {Roux}, {Seabroke}, {Sordo}, {Th{\'e}venin}, {Gracia-Abril}, {Portell}, {Teyssier}, {Altmann}, {Andrae}, {Audard}, {Bellas-Velidis}, {Benson}, {Berthier}, {Blomme}, {Burgess}, {Busonero}, {Busso}, {C{\'a}novas}, {Carry}, {Cellino}, {Cheek}, {Clementini}, {Damerdji}, {Davidson}, {de Teodoro}, {Nu{\~n}ez Campos}, {Delchambre}, {Dell'Oro}, {Esquej},
  {Fern{\'a}ndez-Hern{\'a}ndez}, {Fraile}, {Garabato}, {Garc{\'\i}a-Lario}, {Gosset}, {Haigron}, {Halbwachs}, {Hambly}, {Harrison}, {Hern{\'a}ndez}, {Hestroffer}, {Hodgkin}, {Holl}, {Jan{\ss}en}, {Jevardat de Fombelle}, {Jordan}, {Krone-Martins}, {Lanzafame}, {L{\"o}ffler}, {Marchal}, {Marrese}, {Moitinho}, {Muinonen}, {Osborne}, {Pancino}, {Pauwels}, {Recio-Blanco}, {Reyl{\'e}}, {Riello}, {Rimoldini}, {Roegiers}, {Rybizki}, {Sarro}, {Siopis}, {Smith}, {Sozzetti}, {Utrilla}, {van Leeuwen}, {Abbas}, {{\'A}brah{\'a}m}, {Abreu Aramburu}, {Aerts}, {Aguado}, {Ajaj}, {Aldea-Montero}, {Altavilla}, {{\'A}lvarez}, {Alves}, {Anders}, {Anderson}, {Anglada Varela}, {Antoja}, {Baines}, {Baker}, {Balaguer-N{\'u}{\~n}ez}, {Balbinot}, {Balog}, {Barache}, {Barbato}, {Barros}, {Barstow}, {Bartolom{\'e}}, {Bassilana}, {Bauchet}, {Becciani}, {Bellazzini}, {Berihuete}, {Bernet}, {Bertone}, {Bianchi}, {Binnenfeld}, {Blanco-Cuaresma}, {Blazere}, {Boch}, {Bombrun}, {Bossini}, {Bouquillon}, {Bragaglia}, {Bramante}, {Breedt},
  {Bressan}, {Brouillet}, {Brugaletta}, {Bucciarelli}, {Burlacu}, {Butkevich}, {Buzzi}, {Caffau}, {Cancelliere}, {Cantat-Gaudin}, {Carballo}, {Carlucci}, {Carnerero}, {Carrasco}, {Casamiquela}, {Castellani}, {Castro-Ginard}, {Chaoul}, {Charlot}, {Chemin}, {Chiaramida}, {Chiavassa}, {Chornay}, {Comoretto}, {Contursi}, {Cooper}, {Cornez}, {Cowell}, {Crifo}, {Cropper}, {Crosta}, {Crowley}, {Dafonte}, {Dapergolas}, {David}, {David}, {de Laverny}, {De Luise}, {De March}, {De Ridder}, {de Souza}, {de Torres}, {del Peloso}, {del Pozo}, {Delbo}, {Delgado}, {Delisle}, {Demouchy}, {Dharmawardena}, {Di Matteo}, {Diakite}, {Diener}, {Distefano}, {Dolding}, {Edvardsson}, {Enke}, {Fabre}, {Fabrizio}, {Faigler}, {Fedorets}, {Fernique}, {Fienga}, {Figueras}, {Fournier}, {Fouron}, {Fragkoudi}, {Gai}, {Garcia-Gutierrez}, {Garcia-Reinaldos}, {Garc{\'\i}a-Torres}, {Garofalo}, {Gavel}, {Gavras}, {Gerlach}, {Geyer}, {Giacobbe}, {Gilmore}, {Girona}, {Giuffrida}, {Gomel}, {Gomez}, {Gonz{\'a}lez-N{\'u}{\~n}ez},
  {Gonz{\'a}lez-Santamar{\'\i}a}, {Gonz{\'a}lez-Vidal}, {Granvik}, {Guillout}, {Guiraud}, {Guti{\'e}rrez-S{\'a}nchez}, {Guy}, {Hatzidimitriou}, {Hauser}, {Haywood}, {Helmer}, {Helmi}, {Sarmiento}, {Hidalgo}, {Hilger}, {H{\l}adczuk}, {Hobbs}, {Holland}, {Huckle}, {Jardine}, {Jasniewicz}, {Jean-Antoine Piccolo}, {Jim{\'e}nez-Arranz}, {Jorissen}, {Juaristi Campillo}, {Julbe}, {Karbevska}, {Kervella}, {Khanna}, {Kontizas}, {Kordopatis}, {Korn}, {K{\'o}sp{\'a}l}, {Kostrzewa-Rutkowska}, {Kruszy{\'n}ska}, {Kun}, {Laizeau}, {Lambert}, {Lanza}, {Lasne}, {Le Campion}, {Lebreton}, {Lebzelter}, {Leccia}, {Leclerc}, {Lecoeur-Taibi}, {Liao}, {Licata}, {Lindstr{\o}m}, {Lister}, {Livanou}, {Lobel}, {Lorca}, {Loup}, {Madrero Pardo}, {Magdaleno Romeo}, {Managau}, {Mann}, {Manteiga}, {Marchant}, {Marconi}, {Marcos}, {Marcos Santos}, {Mar{\'\i}n Pina}, {Marinoni}, {Marocco}, {Marshall}, {Martin Polo}, {Mart{\'\i}n-Fleitas}, {Marton}, {Mary}, {Masip}, {Massari}, {Mastrobuono-Battisti}, {Mazeh}, {McMillan}, {Messina}, {Michalik},
  {Millar}, {Mints}, {Molina}, {Molinaro}, {Moln{\'a}r}, {Monari}, {Mongui{\'o}}, {Montegriffo}, {Montero}, {Mor}, {Mora}, {Morbidelli}, {Morel}, {Morris}, {Muraveva}, {Murphy}, {Musella}, {Nagy}, {Noval}, {Oca{\~n}a}, {Ogden}, {Ordenovic}, {Osinde}, {Pagani}, {Pagano}, {Palaversa}, {Palicio}, {Pallas-Quintela}, {Panahi}, {Payne-Wardenaar}, {Pe{\~n}alosa Esteller}, {Penttil{\"a}}, {Pichon}, {Piersimoni}, {Pineau}, {Plachy}, {Plum}, {Poggio}, {Pr{\v{s}}a}, {Pulone}, {Racero}, {Ragaini}, {Rainer}, {Raiteri}, {Rambaux}, {Ramos}, {Ramos-Lerate}, {Re Fiorentin}, {Regibo}, {Richards}, {Rios Diaz}, {Ripepi}, {Riva}, {Rix}, {Rixon}, {Robichon}, {Robin}, {Robin}, {Roelens}, {Rogues}, {Rohrbasser}, {Romero-G{\'o}mez}, {Rowell}, {Royer}, {Ruz Mieres}, {Rybicki}, {Sadowski}, {S{\'a}ez N{\'u}{\~n}ez}, {Sagrist{\`a} Sell{\'e}s}, {Sahlmann}, {Salguero}, {Samaras}, {Sanchez Gimenez}, {Sanna}, {Santove{\~n}a}, {Sarasso}, {Schultheis}, {Sciacca}, {Segol}, {Segovia}, {S{\'e}gransan}, {Semeux}, {Shahaf}, {Siddiqui}, {Siebert},
  {Siltala}, {Silvelo}, {Slezak}, {Slezak}, {Smart}, {Snaith}, {Solano}, {Solitro}, {Souami}, {Souchay}, {Spagna}, {Spina}, {Spoto}, {Steele}, {Steidelm{\"u}ller}, {Stephenson}, {S{\"u}veges}, {Surdej}, {Szabados}, {Szegedi-Elek}, {Taris}, {Taylor}, {Teixeira}, {Tolomei}, {Tonello}, {Torra}, {Torra}, {Torralba Elipe}, {Trabucchi}, {Tsounis}, {Turon}, {Ulla}, {Unger}, {Vaillant}, {van Dillen}, {van Reeven}, {Vanel}, {Vecchiato}, {Viala}, {Vicente}, {Voutsinas}, {Weiler}, {Wevers}, {Wyrzykowski}, {Yoldas}, {Yvard}, {Zhao}, {Zorec}, {Zucker}, \& {Zwitter}}]{gaia2023}
{Gaia Collaboration}, {Vallenari}, A., {Brown}, A.~G.~A., {et~al.} 2023, \aap, 674, A1, \dodoi{10.1051/0004-6361/202243940}

\bibitem[{{Garc{\'\i}a P{\'e}rez} {et~al.}(2016){Garc{\'\i}a P{\'e}rez}, {Allende Prieto}, {Holtzman}, {Shetrone}, {M{\'e}sz{\'a}ros}, {Bizyaev}, {Carrera}, {Cunha}, {Garc{\'\i}a-Hern{\'a}ndez}, {Johnson}, {Majewski}, {Nidever}, {Schiavon}, {Shane}, {Smith}, {Sobeck}, {Troup}, {Zamora}, {Weinberg}, {Bovy}, {Eisenstein}, {Feuillet}, {Frinchaboy}, {Hayden}, {Hearty}, {Nguyen}, {O'Connell}, {Pinsonneault}, {Wilson}, \& {Zasowski}}]{garcia2016}
{Garc{\'\i}a P{\'e}rez}, A.~E., {Allende Prieto}, C., {Holtzman}, J.~A., {et~al.} 2016, \aj, 151, 144, \dodoi{10.3847/0004-6256/151/6/144}

\bibitem[{{GRAVITY Collaboration} {et~al.}(2021){GRAVITY Collaboration}, {Abuter}, {Amorim}, {Baub{\"o}ck}, {Berger}, {Bonnet}, {Brandner}, {Cl{\'e}net}, {Davies}, {de Zeeuw}, {Dexter}, {Dallilar}, {Drescher}, {Eckart}, {Eisenhauer}, {F{\"o}rster Schreiber}, {Garcia}, {Gao}, {Gendron}, {Genzel}, {Gillessen}, {Habibi}, {Haubois}, {Hei{\ss}el}, {Henning}, {Hippler}, {Horrobin}, {Jim{\'e}nez-Rosales}, {Jochum}, {Jocou}, {Kaufer}, {Kervella}, {Lacour}, {Lapeyr{\`e}re}, {Le Bouquin}, {L{\'e}na}, {Lutz}, {Nowak}, {Ott}, {Paumard}, {Perraut}, {Perrin}, {Pfuhl}, {Rabien}, {Rodr{\'\i}guez-Coira}, {Shangguan}, {Shimizu}, {Scheithauer}, {Stadler}, {Straub}, {Straubmeier}, {Sturm}, {Tacconi}, {Vincent}, {von Fellenberg}, {Waisberg}, {Widmann}, {Wieprecht}, {Wiezorrek}, {Woillez}, {Yazici}, {Young}, \& {Zins}}]{Gravity:2021}
{GRAVITY Collaboration}, {Abuter}, R., {Amorim}, A., {et~al.} 2021, \aap, 647, A59, \dodoi{10.1051/0004-6361/202040208}

\bibitem[{{Grevesse} {et~al.}(2007){Grevesse}, {Asplund}, \& {Sauval}}]{grevesse2007}
{Grevesse}, N., {Asplund}, M., \& {Sauval}, A.~J. 2007, \ssr, 130, 105, \dodoi{10.1007/s11214-007-9173-7}

\bibitem[{{Grieco} {et~al.}(2012){Grieco}, {Matteucci}, {Pipino}, \& {Cescutti}}]{grieco2012}
{Grieco}, V., {Matteucci}, F., {Pipino}, A., \& {Cescutti}, G. 2012, \aap, 548, A60, \dodoi{10.1051/0004-6361/201219761}

\bibitem[{{Grieco} {et~al.}(2015){Grieco}, {Matteucci}, {Ryde}, {Schultheis}, \& {Uttenthaler}}]{grieco2015}
{Grieco}, V., {Matteucci}, F., {Ryde}, N., {Schultheis}, M., \& {Uttenthaler}, S. 2015, \mnras, 450, 2094, \dodoi{10.1093/mnras/stv729}

\bibitem[{{Griffith} {et~al.}(2019){Griffith}, {Johnson}, \& {Weinberg}}]{griffith2019}
{Griffith}, E., {Johnson}, J.~A., \& {Weinberg}, D.~H. 2019, \apj, 886, 84, \dodoi{10.3847/1538-4357/ab4b5d}

\bibitem[{{Griffith} {et~al.}(2021){Griffith}, {Weinberg}, {Johnson}, {Beaton}, {Garc{\'\i}a-Hern{\'a}ndez}, {Hasselquist}, {Holtzman}, {Johnson}, {J{\"o}nsson}, {Lane}, {Nataf}, \& {Roman-Lopes}}]{griffith2021a}
{Griffith}, E., {Weinberg}, D.~H., {Johnson}, J.~A., {et~al.} 2021, \apj, 909, 77, \dodoi{10.3847/1538-4357/abd6be}

\bibitem[{{Griffith} {et~al.}(2024){Griffith}, {Hogg}, {Dalcanton}, {Hasselquist}, {Ratcliffe}, {Ness}, \& {Weinberg}}]{griffith2024}
{Griffith}, E.~J., {Hogg}, D.~W., {Dalcanton}, J.~J., {et~al.} 2024, \aj, 167, 98, \dodoi{10.3847/1538-3881/ad19c7}, (G24)

\bibitem[{{Griffith} {et~al.}(2022){Griffith}, {Weinberg}, {Buder}, {Johnson}, {Johnson}, \& {Vincenzo}}]{griffith2022}
{Griffith}, E.~J., {Weinberg}, D.~H., {Buder}, S., {et~al.} 2022, \apj, 931, 23, \dodoi{10.3847/1538-4357/ac5826}

\bibitem[{{Gronow} {et~al.}(2021){Gronow}, {C{\^o}t{\'e}}, {Lach}, {Seitenzahl}, {Collins}, {Sim}, \& {R{\"o}pke}}]{gronow2021}
{Gronow}, S., {C{\^o}t{\'e}}, B., {Lach}, F., {et~al.} 2021, \aap, 656, A94, \dodoi{10.1051/0004-6361/202140881}

\bibitem[{{Gunn} {et~al.}(2006){Gunn}, {Siegmund}, {Mannery}, {Owen}, {Hull}, {Leger}, {Carey}, {Knapp}, {York}, {Boroski}, {Kent}, {Lupton}, {Rockosi}, {Evans}, {Waddell}, {Anderson}, {Annis}, {Barentine}, {Bartoszek}, {Bastian}, {Bracker}, {Brewington}, {Briegel}, {Brinkmann}, {Brown}, {Carr}, {Czarapata}, {Drennan}, {Dombeck}, {Federwitz}, {Gillespie}, {Gonzales}, {Hansen}, {Harvanek}, {Hayes}, {Jordan}, {Kinney}, {Klaene}, {Kleinman}, {Kron}, {Kresinski}, {Lee}, {Limmongkol}, {Lindenmeyer}, {Long}, {Loomis}, {McGehee}, {Mantsch}, {Neilsen}, {Neswold}, {Newman}, {Nitta}, {Peoples}, {Pier}, {Prieto}, {Prosapio}, {Rivetta}, {Schneider}, {Snedden}, \& {Wang}}]{gunn2006}
{Gunn}, J.~E., {Siegmund}, W.~A., {Mannery}, E.~J., {et~al.} 2006, \aj, 131, 2332, \dodoi{10.1086/500975}

\bibitem[{{Hackshaw} {et~al.}(2024){Hackshaw}, {Hawkins}, {Filion}, {Horta}, {Laporte}, {Carr}, \& {Price-Whelan}}]{hackshaw2024}
{Hackshaw}, Z., {Hawkins}, K., {Filion}, C., {et~al.} 2024, arXiv e-prints, arXiv:2405.18120, \dodoi{10.48550/arXiv.2405.18120}

\bibitem[{{Harris} {et~al.}(2020){Harris}, {Millman}, {van der Walt}, {Gommers}, {Virtanen}, {Cournapeau}, {Wieser}, {Taylor}, {Berg}, {Smith}, {Kern}, {Picus}, {Hoyer}, {van Kerkwijk}, {Brett}, {Haldane}, {del R{\'\i}o}, {Wiebe}, {Peterson}, {G{\'e}rard-Marchant}, {Sheppard}, {Reddy}, {Weckesser}, {Abbasi}, {Gohlke}, \& {Oliphant}}]{harris2020}
{Harris}, C.~R., {Millman}, K.~J., {van der Walt}, S.~J., {et~al.} 2020, \nat, 585, 357, \dodoi{10.1038/s41586-020-2649-2}

\bibitem[{{Hasselquist} {et~al.}(2024){Hasselquist}, {Hayes}, {Griffith}, {Weinberg}, {Sit}, {Beaton}, \& {Horta}}]{hasselquist2024}
{Hasselquist}, S., {Hayes}, C.~R., {Griffith}, E.~J., {et~al.} 2024, arXiv e-prints, arXiv:2408.10393, \dodoi{10.48550/arXiv.2408.10393}

\bibitem[{{Hawkins}(2023)}]{hawkins2023}
{Hawkins}, K. 2023, \mnras, 525, 3318, \dodoi{10.1093/mnras/stad1244}

\bibitem[{{Hayden} {et~al.}(2017){Hayden}, {Recio-Blanco}, {de Laverny}, {Mikolaitis}, \& {Worley}}]{hayden2017}
{Hayden}, M.~R., {Recio-Blanco}, A., {de Laverny}, P., {Mikolaitis}, S., \& {Worley}, C.~C. 2017, \aap, 608, L1, \dodoi{10.1051/0004-6361/201731494}

\bibitem[{{Hayden} {et~al.}(2014){Hayden}, {Holtzman}, {Bovy}, {Majewski}, {Johnson}, {Allende Prieto}, {Beers}, {Cunha}, {Frinchaboy}, {Garc{\'\i}a P{\'e}rez}, {Girardi}, {Hearty}, {Lee}, {Nidever}, {Schiavon}, {Schlesinger}, {Schneider}, {Schultheis}, {Shetrone}, {Smith}, {Zasowski}, {Bizyaev}, {Feuillet}, {Hasselquist}, {Kinemuchi}, {Malanushenko}, {Malanushenko}, {O'Connell}, {Pan}, \& {Stassun}}]{hayden2014}
{Hayden}, M.~R., {Holtzman}, J.~A., {Bovy}, J., {et~al.} 2014, \aj, 147, 116, \dodoi{10.1088/0004-6256/147/5/116}

\bibitem[{{Hayden} {et~al.}(2015){Hayden}, {Bovy}, {Holtzman}, {Nidever}, {Bird}, {Weinberg}, {Andrews}, {Majewski}, {Allende Prieto}, {Anders}, {Beers}, {Bizyaev}, {Chiappini}, {Cunha}, {Frinchaboy}, {Garc{\'\i}a-Her{\'n}andez}, {Garc{\'\i}a P{\'e}rez}, {Girardi}, {Harding}, {Hearty}, {Johnson}, {M{\'e}sz{\'a}ros}, {Minchev}, {O'Connell}, {Pan}, {Robin}, {Schiavon}, {Schneider}, {Schultheis}, {Shetrone}, {Skrutskie}, {Steinmetz}, {Smith}, {Wilson}, {Zamora}, \& {Zasowski}}]{hayden2015}
{Hayden}, M.~R., {Bovy}, J., {Holtzman}, J.~A., {et~al.} 2015, \apj, 808, 132, \dodoi{10.1088/0004-637X/808/2/132}

\bibitem[{{Hayden} {et~al.}(2022){Hayden}, {Sharma}, {Bland-Hawthorn}, {Spina}, {Buder}, {Ciuc{\u{a}}}, {Asplund}, {Casey}, {De Silva}, {D'Orazi}, {Freeman}, {Kos}, {Lewis}, {Lin}, {Lind}, {Martell}, {Schlesinger}, {Simpson}, {Zucker}, {Zwitter}, {Chen}, {{\v{C}}otar}, {Feuillet}, {Horner}, {Joyce}, {Nordlander}, {Stello}, {Tepper-Garcia}, {Ting}, {Wang}, {Wittenmyer}, \& {Wyse}}]{hayden2022}
{Hayden}, M.~R., {Sharma}, S., {Bland-Hawthorn}, J., {et~al.} 2022, \mnras, 517, 5325, \dodoi{10.1093/mnras/stac2787}

\bibitem[{{Hayes} {et~al.}(2018){Hayes}, {Majewski}, {Shetrone}, {Fern{\'a}ndez-Alvar}, {Allende Prieto}, {Schuster}, {Carigi}, {Cunha}, {Smith}, {Sobeck}, {Almeida}, {Beers}, {Carrera}, {Fern{\'a}ndez-Trincado}, {Garc{\'\i}a-Hern{\'a}ndez}, {Geisler}, {Lane}, {Lucatello}, {Matthews}, {Minniti}, {Nitschelm}, {Tang}, {Tissera}, \& {Zamora}}]{hayes2018}
{Hayes}, C.~R., {Majewski}, S.~R., {Shetrone}, M., {et~al.} 2018, \apj, 852, 49, \dodoi{10.3847/1538-4357/aa9cec}

\bibitem[{{Helmi} {et~al.}(2018){Helmi}, {Babusiaux}, {Koppelman}, {Massari}, {Veljanoski}, \& {Brown}}]{helmi2018}
{Helmi}, A., {Babusiaux}, C., {Koppelman}, H.~H., {et~al.} 2018, \nat, 563, 85, \dodoi{10.1038/s41586-018-0625-x}

\bibitem[{{Holtzman} {et~al.}(2015){Holtzman}, {Shetrone}, {Johnson}, {Allende Prieto}, {Anders}, {Andrews}, {Beers}, {Bizyaev}, {Blanton}, {Bovy}, {Carrera}, {Chojnowski}, {Cunha}, {Eisenstein}, {Feuillet}, {Frinchaboy}, {Galbraith-Frew}, {Garc{\'\i}a P{\'e}rez}, {Garc{\'\i}a-Hern{\'a}ndez}, {Hasselquist}, {Hayden}, {Hearty}, {Ivans}, {Majewski}, {Martell}, {Meszaros}, {Muna}, {Nidever}, {Nguyen}, {O'Connell}, {Pan}, {Pinsonneault}, {Robin}, {Schiavon}, {Shane}, {Sobeck}, {Smith}, {Troup}, {Weinberg}, {Wilson}, {Wood-Vasey}, {Zamora}, \& {Zasowski}}]{holtzman2015}
{Holtzman}, J.~A., {Shetrone}, M., {Johnson}, J.~A., {et~al.} 2015, \aj, 150, 148, \dodoi{10.1088/0004-6256/150/5/148}

\bibitem[{{Horta} {et~al.}(2021){Horta}, {Schiavon}, {Mackereth}, {Pfeffer}, {Mason}, {Kisku}, {Fragkoudi}, {Allende Prieto}, {Cunha}, {Hasselquist}, {Holtzman}, {Majewski}, {Nataf}, {O'Connell}, {Schultheis}, \& {Smith}}]{horta2021}
{Horta}, D., {Schiavon}, R.~P., {Mackereth}, J.~T., {et~al.} 2021, \mnras, 500, 1385, \dodoi{10.1093/mnras/staa2987}

\bibitem[{{Horta} {et~al.}(2022){Horta}, {Schiavon}, {Mackereth}, {Weinberg}, {Hasselquist}, {Feuillet}, {O'Connell}, {Anguiano}, {Allende-Prieto}, {Beaton}, {Bizyaev}, {Cunha}, {Geisler}, {Garc{\'\i}a-Hern{\'a}ndez}, {Holtzman}, {J{\"o}nsson}, {Lane}, {Majewski}, {M{\'e}sz{\'a}ros}, {Minniti}, {Nitschelm}, {Shetrone}, {Smith}, \& {Zasowski}}]{horta2022}
---. 2022, arXiv e-prints, arXiv:2204.04233.
\newblock \doarXiv{2204.04233}

\bibitem[{{Horta} {et~al.}(2023){Horta}, {Schiavon}, {Mackereth}, {Weinberg}, {Hasselquist}, {Feuillet}, {O'Connell}, {Anguiano}, {Allende-Prieto}, {Beaton}, {Bizyaev}, {Cunha}, {Geisler}, {Garc{\'\i}a-Hern{\'a}ndez}, {Holtzman}, {J{\"o}nsson}, {Lane}, {Majewski}, {M{\'e}sz{\'a}ros}, {Minniti}, {Nitschelm}, {Shetrone}, {Smith}, \& {Zasowski}}]{horta2023}
---. 2023, \mnras, 520, 5671, \dodoi{10.1093/mnras/stac3179}

\bibitem[{{Hubeny} {et~al.}(2021){Hubeny}, {Allende Prieto}, {Osorio}, \& {Lanz}}]{hubeny2021}
{Hubeny}, I., {Allende Prieto}, C., {Osorio}, Y., \& {Lanz}, T. 2021, arXiv e-prints, arXiv:2104.02829, \dodoi{10.48550/arXiv.2104.02829}

\bibitem[{{Hunt} {et~al.}(2022){Hunt}, {Price-Whelan}, {Johnston}, \& {Darragh-Ford}}]{Hunt:2022}
{Hunt}, J. A.~S., {Price-Whelan}, A.~M., {Johnston}, K.~V., \& {Darragh-Ford}, E. 2022, \mnras, 516, L7, \dodoi{10.1093/mnrasl/slac082}

\bibitem[{{Hunter}(2007)}]{hunter2007}
{Hunter}, J.~D. 2007, Computing in Science and Engineering, 9, 90, \dodoi{10.1109/MCSE.2007.55}

\bibitem[{{Imig} {et~al.}(2023){Imig}, {Price}, {Holtzman}, {Stone-Martinez}, {Majewski}, {Weinberg}, {Johnson}, {Allende Prieto}, {Beaton}, {Beers}, {Bizyaev}, {Blanton}, {Brownstein}, {Cunha}, {Fern{\'a}ndez-Trincado}, {Feuillet}, {Hasselquist}, {Hayes}, {J{\"o}nsson}, {Lane}, {Lian}, {M{\'e}sz{\'a}ros}, {Nidever}, {Robin}, {Shetrone}, {Smith}, \& {Wilson}}]{imig2023}
{Imig}, J., {Price}, C., {Holtzman}, J.~A., {et~al.} 2023, \apj, 954, 124, \dodoi{10.3847/1538-4357/ace9b8}

\bibitem[{{Johnson} \& {Weinberg}(2020)}]{johnson2020}
{Johnson}, J.~W., \& {Weinberg}, D.~H. 2020, \mnras, 498, 1364, \dodoi{10.1093/mnras/staa2431}

\bibitem[{{Johnson} {et~al.}(2023){Johnson}, {Weinberg}, {Vincenzo}, {Bird}, \& {Griffith}}]{johnson2023}
{Johnson}, J.~W., {Weinberg}, D.~H., {Vincenzo}, F., {Bird}, J.~C., \& {Griffith}, E.~J. 2023, \mnras, 520, 782, \dodoi{10.1093/mnras/stad057}

\bibitem[{{Johnson} {et~al.}(2024){Johnson}, {Weinberg}, {Blanc}, {Bonaca}, {Rudie}, {Yuxi}, {Lu}, {Reichardt Chu}, {Griffith}, {Sit}, {Johnson}, {Dubay}, {Weller}, {Boyea}, \& {Bird}}]{johnson2024}
{Johnson}, J.~W., {Weinberg}, D.~H., {Blanc}, G.~A., {et~al.} 2024, arXiv e-prints, arXiv:2410.13256, \dodoi{10.48550/arXiv.2410.13256}

\bibitem[{{J{\"o}nsson} {et~al.}(2020){J{\"o}nsson}, {Holtzman}, {Allende Prieto}, {Cunha}, {Garc{\'\i}a-Hern{\'a}ndez}, {Hasselquist}, {Masseron}, {Osorio}, {Shetrone}, {Smith}, {Stringfellow}, {Bizyaev}, {Edvardsson}, {Majewski}, {M{\'e}sz{\'a}ros}, {Souto}, {Zamora}, {Beaton}, {Bovy}, {Donor}, {Pinsonneault}, {Poovelil}, \& {Sobeck}}]{jonsson2020}
{J{\"o}nsson}, H., {Holtzman}, J.~A., {Allende Prieto}, C., {et~al.} 2020, \aj, 160, 120, \dodoi{10.3847/1538-3881/aba592}

\bibitem[{{Karakas} \& {Lugaro}(2016)}]{karakas2016}
{Karakas}, A.~I., \& {Lugaro}, M. 2016, \apj, 825, 26, \dodoi{10.3847/0004-637X/825/1/26}

\bibitem[{{Kollmeier} {et~al.}(2019){Kollmeier}, {Anderson}, {Blanc}, {Blanton}, {Covey}, {Crane}, {Drory}, {Frinchaboy}, {Froning}, {Johnson}, {Kneib}, {Kreckel}, {Merloni}, {Pellegrini}, {Pogge}, {Ramirez}, {Rix}, {Sayres}, {S{\'a}nchez-Gallego}, {Shen}, {Tkachenko}, {Trump}, {Tuttle}, {Weijmans}, {Zasowski}, {Barbuy}, {Beaton}, {Bergemann}, {Bochanski}, {Brandt}, {Casey}, {Cherinka}, {Eracleous}, {Fan}, {Garc{\'\i}a}, {Green}, {Hekker}, {Lane}, {Longa-Pe{\~n}a}, {Mathur}, {Meza}, {Minchev}, {Myers}, {Nidever}, {Nitschelm}, {O'Connell}, {Price-Whelan}, {Raddick}, {Rossi}, {Sankrit}, {Simon}, {Stutz}, {Ting}, {Trakhtenbrot}, {Weaver}, {Willmer}, \& {Weinberg}}]{kollmeier2019}
{Kollmeier}, J., {Anderson}, S.~F., {Blanc}, G.~A., {et~al.} 2019, in Bulletin of the American Astronomical Society, Vol.~51, 274

\bibitem[{{Mackereth} {et~al.}(2017){Mackereth}, {Bovy}, {Schiavon}, {Zasowski}, {Cunha}, {Frinchaboy}, {Garc{\'\i}a Perez}, {Hayden}, {Holtzman}, {Majewski}, {M{\'e}sz{\'a}ros}, {Nidever}, {Pinsonneault}, \& {Shetrone}}]{mackereth2017}
{Mackereth}, J.~T., {Bovy}, J., {Schiavon}, R.~P., {et~al.} 2017, \mnras, 471, 3057, \dodoi{10.1093/mnras/stx1774}

\bibitem[{{Mackereth} {et~al.}(2019){Mackereth}, {Schiavon}, {Pfeffer}, {Hayes}, {Bovy}, {Anguiano}, {Allende Prieto}, {Hasselquist}, {Holtzman}, {Johnson}, {Majewski}, {O'Connell}, {Shetrone}, {Tissera}, \& {Fern{\'a}ndez-Trincado}}]{mackereth2019}
{Mackereth}, J.~T., {Schiavon}, R.~P., {Pfeffer}, J., {et~al.} 2019, \mnras, 482, 3426, \dodoi{10.1093/mnras/sty2955}

\bibitem[{{Majewski} {et~al.}(2017){Majewski}, {Schiavon}, {Frinchaboy}, {Allende Prieto}, {Barkhouser}, {Bizyaev}, {Blank}, {Brunner}, {Burton}, {Carrera}, {Chojnowski}, {Cunha}, {Epstein}, {Fitzgerald}, {Garc{\'\i}a P{\'e}rez}, {Hearty}, {Henderson}, {Holtzman}, {Johnson}, {Lam}, {Lawler}, {Maseman}, {M{\'e}sz{\'a}ros}, {Nelson}, {Nguyen}, {Nidever}, {Pinsonneault}, {Shetrone}, {Smee}, {Smith}, {Stolberg}, {Skrutskie}, {Walker}, {Wilson}, {Zasowski}, {Anders}, {Basu}, {Beland}, {Blanton}, {Bovy}, {Brownstein}, {Carlberg}, {Chaplin}, {Chiappini}, {Eisenstein}, {Elsworth}, {Feuillet}, {Fleming}, {Galbraith-Frew}, {Garc{\'\i}a}, {Garc{\'\i}a-Hern{\'a}ndez}, {Gillespie}, {Girardi}, {Gunn}, {Hasselquist}, {Hayden}, {Hekker}, {Ivans}, {Kinemuchi}, {Klaene}, {Mahadevan}, {Mathur}, {Mosser}, {Muna}, {Munn}, {Nichol}, {O'Connell}, {Parejko}, {Robin}, {Rocha-Pinto}, {Schultheis}, {Serenelli}, {Shane}, {Silva Aguirre}, {Sobeck}, {Thompson}, {Troup}, {Weinberg}, \& {Zamora}}]{majewski2017}
{Majewski}, S.~R., {Schiavon}, R.~P., {Frinchaboy}, P.~M., {et~al.} 2017, \aj, 154, 94, \dodoi{10.3847/1538-3881/aa784d}

\bibitem[{{Martig} {et~al.}(2016){Martig}, {Minchev}, {Ness}, {Fouesneau}, \& {Rix}}]{martig2016}
{Martig}, M., {Minchev}, I., {Ness}, M., {Fouesneau}, M., \& {Rix}, H.-W. 2016, \apj, 831, 139, \dodoi{10.3847/0004-637X/831/2/139}

\bibitem[{{McKinnon} {et~al.}(2024){McKinnon}, {Ness}, {Rockosi}, \& {Guhathakurta}}]{mickinnon2024}
{McKinnon}, K.~A., {Ness}, M.~K., {Rockosi}, C.~M., \& {Guhathakurta}, P. 2024, \apj, 965, 120, \dodoi{10.3847/1538-4357/ad2859}

\bibitem[{{Micali} {et~al.}(2013){Micali}, {Matteucci}, \& {Romano}}]{micali2018}
{Micali}, A., {Matteucci}, F., \& {Romano}, D. 2013, \mnras, 436, 1648, \dodoi{10.1093/mnras/stt1681}

\bibitem[{{Minchev} {et~al.}(2013){Minchev}, {Chiappini}, \& {Martig}}]{minchev2013}
{Minchev}, I., {Chiappini}, C., \& {Martig}, M. 2013, \aap, 558, A9, \dodoi{10.1051/0004-6361/201220189}

\bibitem[{{Ness} {et~al.}(2022){Ness}, {Wheeler}, {McKinnon}, {Horta}, {Casey}, {Cunningham}, \& {Price-Whelan}}]{ness2022}
{Ness}, M.~K., {Wheeler}, A.~J., {McKinnon}, K., {et~al.} 2022, \apj, 926, 144, \dodoi{10.3847/1538-4357/ac4754}

\bibitem[{{Nidever} {et~al.}(2015){Nidever}, {Holtzman}, {Allende Prieto}, {Beland}, {Bender}, {Bizyaev}, {Burton}, {Desphande}, {Fleming}, {Garc{\'\i}a P{\'e}rez}, {Hearty}, {Majewski}, {M{\'e}sz{\'a}ros}, {Muna}, {Nguyen}, {Schiavon}, {Shetrone}, {Skrutskie}, {Sobeck}, \& {Wilson}}]{nidever2015}
{Nidever}, D.~L., {Holtzman}, J.~A., {Allende Prieto}, C., {et~al.} 2015, \aj, 150, 173, \dodoi{10.1088/0004-6256/150/6/173}

\bibitem[{{Nissen} \& {Schuster}(2010)}]{nissen2010}
{Nissen}, P.~E., \& {Schuster}, W.~J. 2010, \aap, 511, L10, \dodoi{10.1051/0004-6361/200913877}

\bibitem[{{Osorio} {et~al.}(2020){Osorio}, {Allende Prieto}, {Hubeny}, {M{\'e}sz{\'a}ros}, \& {Shetrone}}]{osorio2020}
{Osorio}, Y., {Allende Prieto}, C., {Hubeny}, I., {M{\'e}sz{\'a}ros}, S., \& {Shetrone}, M. 2020, \aap, 637, A80, \dodoi{10.1051/0004-6361/201937054}

\bibitem[{pandas~development team(2020)}]{pandasa}
pandas~development team, T. 2020, pandas-dev/pandas: Pandas, latest,  Zenodo, \dodoi{10.5281/zenodo.3509134}

\bibitem[{Pedregosa {et~al.}(2011)Pedregosa, Varoquaux, Gramfort, Michel, Thirion, Grisel, Blondel, Prettenhofer, Weiss, Dubourg, Vanderplas, Passos, Cournapeau, Brucher, Perrot, \& Duchesnay}]{scikit-learn}
Pedregosa, F., Varoquaux, G., Gramfort, A., {et~al.} 2011, Journal of Machine Learning Research, 12, 2825

\bibitem[{{Pinsonneault} {et~al.}(2024){Pinsonneault}, {Zinn}, {Tayar}, {Serenelli}, {Garcia}, {Mathur}, {Vrard}, {Elsworth}, {Mosser}, {Stello}, {Bell}, {Bugnet}, {Corsaro}, {Gaulme}, {Hekker}, {Hon}, {Huber}, {Kallinger}, {Cao}, {Johnson}, {Liagre}, {Patton}, {Santos}, {Basu}, {Beck}, {Beers}, {Chaplin}, {Cunha}, {Frinchaboy}, {Girardi}, {Godoy-Rivera}, {Holtzman}, {Jonsson}, {Meszaros}, {Reyes}, {Rix}, {Shetrone}, {Smith}, {Spoo}, {Stassun}, \& {Wang}}]{pinsonnealt2024}
{Pinsonneault}, M.~H., {Zinn}, J.~C., {Tayar}, J., {et~al.} 2024, arXiv e-prints, arXiv:2410.00102, \dodoi{10.48550/arXiv.2410.00102}

\bibitem[{{Price-Jones} \& {Bovy}(2018)}]{price2018}
{Price-Jones}, N., \& {Bovy}, J. 2018, \mnras, 475, 1410, \dodoi{10.1093/mnras/stx3198}

\bibitem[{{Price-Whelan}(2017)}]{Price-Whelan:2017}
{Price-Whelan}, A.~M. 2017, The Journal of Open Source Software, 2, 388, \dodoi{10.21105/joss.00388}

\bibitem[{{Queiroz} {et~al.}(2023){Queiroz}, {Anders}, {Chiappini}, {Khalatyan}, {Santiago}, {Nepal}, {Steinmetz}, {Gallart}, {Valentini}, {Dal Ponte}, {Barbuy}, {P{\'e}rez-Villegas}, {Masseron}, {Fern{\'a}ndez-Trincado}, {Khoperskov}, {Minchev}, {Fern{\'a}ndez-Alvar}, {Lane}, \& {Nitschelm}}]{starhorse}
{Queiroz}, A.~B.~A., {Anders}, F., {Chiappini}, C., {et~al.} 2023, \aap, 673, A155, \dodoi{10.1051/0004-6361/202245399}

\bibitem[{{Ratcliffe} {et~al.}(2024){Ratcliffe}, {Minchev}, {Cescutti}, {Spitoni}, {J{\"o}nsson}, {Anders}, {Queiroz}, \& {Steinmetz}}]{ratcliffe2024}
{Ratcliffe}, B., {Minchev}, I., {Cescutti}, G., {et~al.} 2024, \mnras, 528, 3464, \dodoi{10.1093/mnras/stae226}

\bibitem[{{Ratcliffe} \& {Ness}(2022)}]{ratcliffe2022}
{Ratcliffe}, B., \& {Ness}, M. 2022, arXiv e-prints, arXiv:2206.02772.
\newblock \doarXiv{2206.02772}

\bibitem[{{Ratcliffe} \& {Ness}(2023)}]{ratcliffe2023}
{Ratcliffe}, B.~L., \& {Ness}, M.~K. 2023, \apj, 943, 92, \dodoi{10.3847/1538-4357/aca8a1}

\bibitem[{{Ratcliffe} {et~al.}(2020){Ratcliffe}, {Ness}, {Johnston}, \& {Sen}}]{ratcliffe2020}
{Ratcliffe}, B.~L., {Ness}, M.~K., {Johnston}, K.~V., \& {Sen}, B. 2020, \apj, 900, 165, \dodoi{10.3847/1538-4357/abac61}

\bibitem[{{Reid} \& {Brunthaler}(2020)}]{Reid:2020}
{Reid}, M.~J., \& {Brunthaler}, A. 2020, \apj, 892, 39, \dodoi{10.3847/1538-4357/ab76cd}

\bibitem[{{Rybizki} {et~al.}(2017){Rybizki}, {Just}, \& {Rix}}]{rybizki2017}
{Rybizki}, J., {Just}, A., \& {Rix}, H.-W. 2017, \aap, 605, A59, \dodoi{10.1051/0004-6361/201730522}

\bibitem[{{Salaris} {et~al.}(2015){Salaris}, {Pietrinferni}, {Piersimoni}, \& {Cassisi}}]{salaris2015}
{Salaris}, M., {Pietrinferni}, A., {Piersimoni}, A.~M., \& {Cassisi}, S. 2015, \aap, 583, A87, \dodoi{10.1051/0004-6361/201526951}

\bibitem[{{Sanders} \& {Binney}(2014)}]{Sanders:2014}
{Sanders}, J.~L., \& {Binney}, J. 2014, \mnras, 441, 3284, \dodoi{10.1093/mnras/stu796}

\bibitem[{{Sanders} \& {Binney}(2016)}]{Sanders:2016}
---. 2016, \mnras, 457, 2107, \dodoi{10.1093/mnras/stw106}

\bibitem[{{Santana} {et~al.}(2021){Santana}, {Beaton}, {Covey}, {O'Connell}, {Longa-Pe{\~n}a}, {Cohen}, {Fern{\'a}ndez-Trincado}, {Hayes}, {Zasowski}, {Sobeck}, {Majewski}, {Chojnowski}, {De Lee}, {Oelkers}, {Stringfellow}, {Almeida}, {Anguiano}, {Donor}, {Frinchaboy}, {Hasselquist}, {Johnson}, {Kollmeier}, {Nidever}, {Price-Whelan}, {Rojas-Arriagada}, {Schultheis}, {Shetrone}, {Simon}, {Aerts}, {Borissova}, {Drout}, {Geisler}, {Law}, {Medina}, {Minniti}, {Monachesi}, {Mu{\~n}oz}, {Poleski}, {Roman-Lopes}, {Schlaufman}, {Stutz}, {Teske}, {Tkachenko}, {Van Saders}, {Weinberger}, \& {Zoccali}}]{santana2021}
{Santana}, F.~A., {Beaton}, R.~L., {Covey}, K.~R., {et~al.} 2021, \aj, 162, 303, \dodoi{10.3847/1538-3881/ac2cbc}

\bibitem[{{Sch{\"o}nrich} \& {Binney}(2009)}]{schonrich2009}
{Sch{\"o}nrich}, R., \& {Binney}, J. 2009, \mnras, 399, 1145, \dodoi{10.1111/j.1365-2966.2009.15365.x}

\bibitem[{{Sharma} {et~al.}(2016){Sharma}, {Stello}, {Bland-Hawthorn}, {Huber}, \& {Bedding}}]{sharma2016}
{Sharma}, S., {Stello}, D., {Bland-Hawthorn}, J., {Huber}, D., \& {Bedding}, T.~R. 2016, \apj, 822, 15, \dodoi{10.3847/0004-637X/822/1/15}

\bibitem[{{Simmerer} {et~al.}(2004){Simmerer}, {Sneden}, {Cowan}, {Collier}, {Woolf}, \& {Lawler}}]{simmerer2004}
{Simmerer}, J., {Sneden}, C., {Cowan}, J.~J., {et~al.} 2004, \apj, 617, 1091, \dodoi{10.1086/424504}

\bibitem[{{Sit} {et~al.}(2024){Sit}, {Weinberg}, {Wheeler}, {Hayes}, {Hasselquist}, {Masseron}, \& {Sobeck}}]{sit2024}
{Sit}, T., {Weinberg}, D.~H., {Wheeler}, A., {et~al.} 2024, arXiv e-prints, arXiv:2403.08067, \dodoi{10.48550/arXiv.2403.08067}, (S24)

\bibitem[{{Spitoni} {et~al.}(2024){Spitoni}, {Matteucci}, {Gratton}, {Ratcliffe}, {Minchev}, \& {Cescutti}}]{spitoni2024}
{Spitoni}, E., {Matteucci}, F., {Gratton}, R., {et~al.} 2024, \aap, 690, A208, \dodoi{10.1051/0004-6361/202450754}

\bibitem[{{Spitoni} {et~al.}(2019){Spitoni}, {Silva Aguirre}, {Matteucci}, {Calura}, \& {Grisoni}}]{spitoni2019}
{Spitoni}, E., {Silva Aguirre}, V., {Matteucci}, F., {Calura}, F., \& {Grisoni}, V. 2019, \aap, 623, A60, \dodoi{10.1051/0004-6361/201834188}

\bibitem[{{Spitoni} {et~al.}(2023){Spitoni}, {Recio-Blanco}, {de Laverny}, {Palicio}, {Kordopatis}, {Schultheis}, {Contursi}, {Poggio}, {Romano}, \& {Matteucci}}]{spitoni2022}
{Spitoni}, E., {Recio-Blanco}, A., {de Laverny}, P., {et~al.} 2023, \aap, 670, A109, \dodoi{10.1051/0004-6361/202244349}

\bibitem[{{Stone-Martinez} {et~al.}(2024){Stone-Martinez}, {Holtzman}, {Imig}, {Nitschelm}, {Stassun}, \& {Brownstein}}]{stone2024}
{Stone-Martinez}, A., {Holtzman}, J.~A., {Imig}, J., {et~al.} 2024, \aj, 167, 73, \dodoi{10.3847/1538-3881/ad12a6}

\bibitem[{{Ting} {et~al.}(2012){Ting}, {Freeman}, {Kobayashi}, {De Silva}, \& {Bland-Hawthorn}}]{ting2012}
{Ting}, Y.-S., {Freeman}, K.~C., {Kobayashi}, C., {De Silva}, G.~M., \& {Bland-Hawthorn}, J. 2012, \mnras, 421, 1231, \dodoi{10.1111/j.1365-2966.2011.20387.x}

\bibitem[{{Ting} \& {Weinberg}(2022)}]{ting2022}
{Ting}, Y.-S., \& {Weinberg}, D.~H. 2022, \apj, 927, 209, \dodoi{10.3847/1538-4357/ac5023}

\bibitem[{{Truran} \& {Arnett}(1971)}]{truran1971}
{Truran}, J.~W., \& {Arnett}, W.~D. 1971, \apss, 11, 430, \dodoi{10.1007/BF00649636}

\bibitem[{{Vincenzo} {et~al.}(2021){Vincenzo}, {Weinberg}, {Montalb{\'a}n}, {Miglio}, {Khan}, {Griffith}, {Hasselquist}, {Johnson}, {Johnson}, {Nitschelm}, \& {Pinsonneault}}]{vincenzo2021b}
{Vincenzo}, F., {Weinberg}, D.~H., {Montalb{\'a}n}, J., {et~al.} 2021, arXiv e-prints, arXiv:2106.03912, \dodoi{10.48550/arXiv.2106.03912}

\bibitem[{{Weinberg} {et~al.}(2019){Weinberg}, {Holtzman}, {Hasselquist}, {Bird}, {Johnson}, {Shetrone}, {Sobeck}, {Allende Prieto}, {Bizyaev}, {Carrera}, {Cohen}, {Cunha}, {Ebelke}, {Fernandez-Trincado}, {Garc{\'\i}a-Hern{\'a}ndez}, {Hayes}, {J{\"o}nsson}, {Lane}, {Majewski}, {Malanushenko}, {M{\'e}sz{\'a}ros}, {Nidever}, {Nitschelm}, {Pan}, {Rix}, {Rybizki}, {Schiavon}, {Schneider}, {Wilson}, \& {Zamora}}]{weinberg2019}
{Weinberg}, D.~H., {Holtzman}, J.~A., {Hasselquist}, S., {et~al.} 2019, \apj, 874, 102, \dodoi{10.3847/1538-4357/ab07c7}

\bibitem[{{Weinberg} {et~al.}(2022){Weinberg}, {Holtzman}, {Johnson}, {Hayes}, {Hasselquist}, {Shetrone}, {Ting}, {Beaton}, {Beers}, {Bird}, {Bizyaev}, {Blanton}, {Cunha}, {Fern{\'a}ndez-Trincado}, {Frinchaboy}, {Garc{\'\i}a-Hern{\'a}ndez}, {Griffith}, {Johnson}, {J{\"o}nsson}, {Lane}, {Leung}, {Mackereth}, {Majewski}, {M{\'e}sz{\'a}ros}, {Nitschelm}, {Pan}, {Schiavon}, {Schneider}, {Schultheis}, {Smith}, {Sobeck}, {Stassun}, {Stringfellow}, {Vincenzo}, {Wilson}, \& {Zasowski}}]{weinberg2022}
{Weinberg}, D.~H., {Holtzman}, J.~A., {Johnson}, J.~A., {et~al.} 2022, \apjs, 260, 32, \dodoi{10.3847/1538-4365/ac6028}

\bibitem[{{W}es {M}c{K}inney(2010)}]{pandasb}
{W}es {M}c{K}inney. 2010, in {P}roceedings of the 9th {P}ython in {S}cience {C}onference, ed. {S}t\'efan van~der {W}alt \& {J}arrod {M}illman, 56 -- 61, \dodoi{10.25080/Majora-92bf1922-00a}

\bibitem[{{Wilson} {et~al.}(2019){Wilson}, {Hearty}, {Skrutskie}, {Majewski}, {Holtzman}, {Eisenstein}, {Gunn}, {Blank}, {Henderson}, {Smee}, {Nelson}, {Nidever}, {Arns}, {Barkhouser}, {Barr}, {Beland}, {Bershady}, {Blanton}, {Brunner}, {Burton}, {Carey}, {Carr}, {Colque}, {Crane}, {Damke}, {Davidson}, {Dean}, {Di Mille}, {Don}, {Ebelke}, {Evans}, {Fitzgerald}, {Gillespie}, {Hall}, {Harding}, {Harding}, {Hammond}, {Hancock}, {Harrison}, {Hope}, {Horne}, {Karakla}, {Lam}, {Leger}, {MacDonald}, {Maseman}, {Matsunari}, {Melton}, {Mitcheltree}, {O'Brien}, {O'Connell}, {Patten}, {Richardson}, {Rieke}, {Rieke}, {Roman-Lopes}, {Schiavon}, {Sobeck}, {Stolberg}, {Stoll}, {Tembe}, {Trujillo}, {Uomoto}, {Vernieri}, {Walker}, {Weinberg}, {Young}, {Anthony-Brumfield}, {Bizyaev}, {Breslauer}, {De Lee}, {Downey}, {Halverson}, {Huehnerhoff}, {Klaene}, {Leon}, {Long}, {Mahadevan}, {Malanushenko}, {Nguyen}, {Owen}, {S{\'a}nchez-Gallego}, {Sayres}, {Shane}, {Shectman}, {Shetrone}, {Skinner}, {Stauffer}, \& {Zhao}}]{wilson2019}
{Wilson}, J.~C., {Hearty}, F.~R., {Skrutskie}, M.~F., {et~al.} 2019, \pasp, 131, 055001, \dodoi{10.1088/1538-3873/ab0075}

\bibitem[{{Xiang} {et~al.}(2019){Xiang}, {Ting}, {Rix}, {Sandford}, {Buder}, {Lind}, {Liu}, {Shi}, \& {Zhang}}]{lamost}
{Xiang}, M., {Ting}, Y.-S., {Rix}, H.-W., {et~al.} 2019, \apjs, 245, 34, \dodoi{10.3847/1538-4365/ab5364}

\bibitem[{{Zasowski} {et~al.}(2013){Zasowski}, {Johnson}, {Frinchaboy}, {Majewski}, {Nidever}, {Rocha Pinto}, {Girardi}, {Andrews}, {Chojnowski}, {Cudworth}, {Jackson}, {Munn}, {Skrutskie}, {Beaton}, {Blake}, {Covey}, {Deshpande}, {Epstein}, {Fabbian}, {Fleming}, {Garcia Hernandez}, {Herrero}, {Mahadevan}, {M{\'e}sz{\'a}ros}, {Schultheis}, {Sellgren}, {Terrien}, {van Saders}, {Allende Prieto}, {Bizyaev}, {Burton}, {Cunha}, {da Costa}, {Hasselquist}, {Hearty}, {Holtzman}, {Garc{\'\i}a P{\'e}rez}, {Maia}, {O'Connell}, {O'Donnell}, {Pinsonneault}, {Santiago}, {Schiavon}, {Shetrone}, {Smith}, \& {Wilson}}]{zasowski2013}
{Zasowski}, G., {Johnson}, J.~A., {Frinchaboy}, P.~M., {et~al.} 2013, \aj, 146, 81, \dodoi{10.1088/0004-6256/146/4/81}

\bibitem[{{Zasowski} {et~al.}(2017){Zasowski}, {Cohen}, {Chojnowski}, {Santana}, {Oelkers}, {Andrews}, {Beaton}, {Bender}, {Bird}, {Bovy}, {Carlberg}, {Covey}, {Cunha}, {Dell'Agli}, {Fleming}, {Frinchaboy}, {Garc{\'\i}a-Hern{\'a}ndez}, {Harding}, {Holtzman}, {Johnson}, {Kollmeier}, {Majewski}, {M{\'e}sz{\'a}ros}, {Munn}, {Mu{\~n}oz}, {Ness}, {Nidever}, {Poleski}, {Rom{\'a}n-Z{\'u}{\~n}iga}, {Shetrone}, {Simon}, {Smith}, {Sobeck}, {Stringfellow}, {Szigeti{\'a}ros}, {Tayar}, \& {Troup}}]{zasowski2017}
{Zasowski}, G., {Cohen}, R.~E., {Chojnowski}, S.~D., {et~al.} 2017, \aj, 154, 198, \dodoi{10.3847/1538-3881/aa8df9}

\end{thebibliography}
\bibliographystyle{aasjournal}

\end{document}